\documentclass[paper]{JHEP3}

\usepackage[dvips]{graphicx}
\usepackage{amsmath,amssymb}
\usepackage{bm}
\usepackage{wasysym}
\newcommand{\uhdb}{{\rm {\tiny M}}}
\newcommand{\dhd}{{\rm {\tiny M}}}

\newcommand {\beq} {\begin{equation}}
\newcommand {\eeq} {\end{equation}}
\newcommand {\beqa}{\begin{eqnarray}}
\newcommand {\eeqa}{\end{eqnarray}}

\title{Covariantized Matrix theory for D-particles}

\author{Tamiaki Yoneya\\ Institute of Physics, The University of Tokyo,\footnote{Emeritus Professor, e-mail: tam at hep1.c.u-tokyo.ac.jp}\,\, 
3-8-1 Komaba, Meguro-ku, Tokyo 153-8902, Japan

School of Graduate Studies, 
The Open University of Japan,\footnote{Visiting Professor}
\,\, 
2-11 Wakaba, Mihama-ku, Chiba 261-8586, Japan
}

\preprint{}	

\abstract{
We reformulate the Matrix theory 
of D-particles in a manifestly Lorentz-covariant fashion in the sense of 
11 dimesnional flat Minkowski space-time, from the viewpoint of the so-called 
DLCQ interpretation of the light-front Matrix theory. 
The theory is characterized by various symmetry properties 
including higher gauge symmetries, 
which contain the usual SU($N$) symmetry as a special case and are extended from the structure 
naturally appearing in association with a discretized version of 
Nambu's 3-bracket. The theory is scale invariant, and 
the emergence of 
the 11 dimensional gravitational length, or M-theory scale, 
is interpreted as a consequence of a 
breaking of the scaling symmetry 
through a super-selection rule. 
In the light-front gauge with 
the DLCQ compactification of 11 dimensions, the theory reduces to 
the usual light-front formulation. In the time-like gauge 
with the ordinary M-theory spatial compactification, it reduces to 
a non-Abelian Born-Infeld-like theory, which in the limit of large $N$ 
becomes equivalent with the original BFSS theory. 
}


\begin{document} 


\section{Introduction}

From the viewpoint of 
exploring non-perturbative formulations of string theory, 
the conjecture of 11 dimensional M-theory 
occupies a special pivotal position in 
providing a candidate for the 
strong-coupling limit of the type IIA (and $E_8\times E_8$ Heterotic) 
string theory. Let us first recall the basic tenets of M-theory 
conjecture:  The background space-time is (10,1) 
space-times instead of (9,1) space-times of string theory. 
The 10-th spatial dimension is compactified, $x^{10}\sim x^{10}+2\pi R_{11}$,  around a circle 
of radius $R_{11}=g_s\ell_s$, with $g_s$ and 
$\ell_s$ being the string coupling of type IIA superstrings and 
fundamental string-length constant, respectively. 
The gravitational scale $\ell_{11}$ in 11 dimensions as the 
sole length scale of M-theory is related to 
these string-theory constants by $\ell_{11}=g_s^{1/3}\ell_s$, so that 
the theory with a finite gravitational length in infinitely ($R_{11}\rightarrow \infty$) extended 11 dimensional space-times  
corresponds to a peculiar limit of string theory characterized by 
$g_s\rightarrow \infty$ and $\ell_s^2=\ell_{11}^3/R_{11}\rightarrow 0$. 
In particular, the gravitational interactions at long distance scales 
much larger than $\ell_{11}$ are expected to be described by the classical theory of
11 dimensional supergravity. Dynamical 
degrees of freedom corresponding to strings 
are expected to be (super) membranes (or M2-branes): 
super membranes wrapped once 
around the compactified circle are supposed to behave as 
fundamental strings 
in the remaining 10 dimensional space-time in the limit $g_s\rightarrow 0$ with 
finite $\ell_s$. 
Various D-brane (and other) excitations of string theory 
also find their roles naturally. For instance, 
D0-branes, namely D-particles, are special Kaluza-Klein excitations 
of 11 dimensional gravitons with 
the {\it single} quantized unit $p_{10}=1/R_{11}$ of momentum along the 
 circle in the 11th dimension. D2-branes are super-membranes lying 
entirely in un-compactified 10 dimensional space-times, 
and D4-branes are wrapped M5-branes which are 5-dimensionally 
extended objects, being dual to M2-branes in the sense of 
electromagnetic duality of Dirac with respect to 
RR gauge fields, and so on. 

In spite of various circumstantial evidence for 
this remarkable conjecture, 
only known and perhaps practically 
workable example of concrete formulations of M-theory 
is the so-called BFSS M(atrix) theory \cite{BFSS}. 
This proposal was originated from a 
coincidence of effective theories 
for two apparently differenct objects, namely, 
D-particles and supermembranes. In the limit of 
small $\ell_s$, the effective low-energy theory \cite{witten}
for many-body dynamics of D-particles is supersymmetric 
SU($N$) Yang-Mills quantum mechanics which is 
obtained from the maximally supersymmetric 
super Yang-Mills theory in 10 dimensions by dimensional reduction 
of the base (9,1) space-time to 
(0,1) world line, in which 9 spatial components of 
gauge fields turn into matrix coordinates as collective 
variables representing motion (diagonal matrix elements) and interaction 
(off-diagonal matrix elements) of 
D-particles in terms of 
short open strings. Essentially the same super 
Yang-Mills quantum mechanics also appears \cite{dhn} as a possible regularization 
of a single super membrane formulated in the light-front 
quantization, approximating to a super membrane 
in an appropriate limit of large $N$. In the latter case, 
the functional space of membrane coordinates defined on 
two-dimensional spatial parameter space of the 
membrane world-volume is replaced by the ring of 
Hermitian $N\times N$ matrices. 
The crux of the proposal was to realize that, 
by uniting these two seemingly different 
interpretations as effective theories,  
the super Yang-Mills matrix model may hopefully provide not only 
a regularization of a single membrane, but more importantly 
would describe also ``partons"  for membranes and in principle 
all other excitations of M-theory in a more fundamental manner. 

Suppose we consider the situation where all of 
constituent partons have a unit 10-th momentum $p_{10}=1/R_{11}$ of the 
same sign (namely, no anti-D-paricles) 
along the compactified circle, the total 10-th momentum of 
a system consisting of $N$ partons is 
 $P_{10}=N/R_{11}=Np_{10}$. In the limit of large 
$N$, it defines an infinite momentum frame $P_{10}
\rightarrow \infty$ along the 
compactifed circle. Then the coincidence between 
the effective {\it non-relativistic} Yang-Mills 
quantum mechanics of D-branes and the light-front 
regularization of supermembrane is understandable. 
Remember the case of a single relativistic particle
with mass-shell condition 
$P^{\mu}P_{\mu}+m^2=0$, 
\begin{equation}
-P^-\equiv P^0-P^{10}=\sqrt{(P^i)^2+m^2+(P^{10})^2}-P^{10}
\rightarrow  \frac{(P^i)^2+m^2}{2P^{10}}
\label{massshell0}
\end{equation}
with the indices $i=1, 2, \ldots, 9$ running only over transverse directions. 
By making identification $P^{10}=N/R_{11}$ for the 
compactified 10-th direction, we expect that this form of $P^0$ 
corresponds to the center-of-mass energy of an 
$N$ D-particle system, providing 
 that $m^2$ is the effective {\it relativistically 
invariant}
squared mass of the system. 
We can also adopt an alternative viewpoint, namely 
the so-called DLCQ (discrete light-cone quantization) interpretation: instead of 10-th spatial 
direction, we can assume \cite{suss} that a light-like direction 
$x^-\equiv x^{10}-x^0$ is compactified into a circle of 
radius $R$ with periodicity $x^-\sim x^-+2\pi R$. 
Then the light-like momentum $P^+\equiv P^{10}+P^0$ is 
discretized, $P^+/2=N/R$. With the same proviso for $N$ again as the size of matrices, we have the same expression as (\ref{massshell0}) 
now as an {\it exact} relation without taking the 
large $N$ limit
\begin{equation}
-P^-=\frac{(P^i)^2+m^2}{P^+}=R\frac{(P^i)^2+m^2}{2N}, 
\end{equation}
but with $R_{11}$ being replaced by $R$. 

The difference of 
these two interpretations lies in the natures of Lorentz symmetry 
in 11 dimensions. 
In the former spatial compactification scheme, a boost 
along the compactified 10-th direction is a discrete change 
of the quantum number $N$ with fixed (and hence Lorentz invariant) $R_{11}$, while in the latter 
that is nothing but 
a continuous rescaling of $R$ with fixed $N$. Thus, in the 
DLCQ interpretation, $N$ is Lorentz-invariant and 
$P^+$ is a continuously varying dynamical variable. In both cases, however,  
the limit of  un-compactification (namely, strong-coupling 
limit of type IIA string theory) requires the large $N$ limit, because  
in the DLCQ case the longitudinal momentum $P^+$ must also become a 
continuous finite variable even in a fixed Lorentz frame which is 
possible only by allowing infinite $R$ and $N$. 
Further arguments \cite{seiberg} justifying the viewpoint of 
the DLCQ interpretation were given, suggesting
 that it could be understood 
as a result of taking a limit of large boost from the former interpretation 
with small spatial compactification radius corresponding to a limit of 
weak string coupling. 
In both cases, 
the parton interpretation of D-particles requires that 
possible KK excitations with multiple units of momenta, such as 
$p_{10}=2/R_{11}$ 
or $p^+=4/R$ and higher, are interpreted as 
composite states of two and higher numbers of partons. 

 It is also to be 
noted that the theory 
naturally describes general 
multi-body states of these composite states, since 
$N\times N$ matrices contain as subsystems block-diagonal 
matrices $N_i\times N_i$ with $N=\sum_i N_i$. 
The off-diagonal blocks then are responsible for 
interactions of these subsystems. 
Therefore, it is essential to treat 
systems with all different $N$'s from $N=2$ to infinity 
on an equal footing, even apart from the requirement 
of including all possible 
values of the total longitudinal momentum. Note also that the exchanges of longitudinal 
momentum $p_{10}$ or $p^+$ among constituent subsystems occur in principle as (non-perturbative) 
processes of rearranging constituent partons in the internal dynamics 
of SU($N$) Yang-Mills (super) quantum mechanics. 

From the late 1990s to the early 2000s, 
numerous works testing the proposal appeared. In particular, 
the DLCQ interpretation made us possible to 
perform certain perturbative analyses of super Yang-Mills 
quantum mechanics in exploring whether 
it gives reasonable gravitational interactions of D-particles and 
other excitations with respect to 
scatterings of those excitations in reduced 10 dimensional 
space-time. Although 
we had various encouraging results supporting the 
M(atrix) theory conjecture, the final conclusion 
has not been reached yet.\footnote{For a nice summary of 
such works, we refer the reader to ref. \cite{taylor} giving 
a reasonably comprehensive review 
of the status with an extensive list of literature until around 2000. Unfortunately, we have not seen much progress since then. One thing among more recent works to be mentioned 
seems that we now have some suggestive results on non-perturbative 
properties using numerical simulations. For instance, 
we have reported results \cite{hanada} about the correlation functions of super 
Yang-Mills quantum mechanics, which are 
consistent with the predictions \cite{SY} obtained from a ``holographic" 
approach on the relation between 
10D reduced 11D supergravity and super Yang-Mills quantum 
mechanics.  }

One of the problems left was whether and how fully Lorentz covariant 
formulations of the theory would be possible. 
If we adopt the viewpoint of the DLCQ interpretation 
supposing that the Matrix theory with finite $N$ already 
gives an exact theory with special light-like compactification, 
it is not unreasonable to believe the existence of 
covariant version of the finite $N$ super Yang-Mills 
mechanics. This is particularly so, if we recall 
that the above relation between the discretized light-like momentum and 
the size of matrices still allows continuously varying $P^+$ with 
an arbitrary (real and positive) parameter $R$ 
corresponding to boost transformations. 
Since $N$ is invariant under boost 
by definition in the DLCQ interpretation, it seems natural to imagine a generalization of 
super Yang-Mills mechanics with full covariance allowing 
general Lorentz transformations for fixed finite $N$ as 
a conserved quantum number, not restricted only to 
boost transformation along the compactified circle, with all of 
the 10+1 directions of eleven dimensional Minkowski space-time being 
treated equally as matrices or some extensions of matrices. Otherwise, it seems 
difficult to justify the DLCQ interpretation. If such a covariant 
theory exists as in the case of the ordinary 
particle mechanics, the DLCQ matrix theory would be 
obtained as an exact theory from a covariantized 
Matrix theory with a Lorentz-invariant effective 
mass square.  Although we have to 
take the limit of large $N$ to elevate it to a full fledged formulation of M-theory, 
a consistent covariant formulation with finite $N$ could be an intermediate step toward our ultimate objective. 

With this motivation in mind, we studied in ref. \cite{almy} the quantization
(or more precisely {\it discretization}) of 
the  Nambu bracket \cite{nambu}. 
The Nambu (-Poisson) bracket naturally appears in covariant 
treatments of classical membranes. For instance, the bosonic action 
of a membrane can be expressed in the form 
\begin{equation}
A_{{\rm mem}}=-\frac{1}{\ell_{11}^3}
 \int d^3\xi \Bigl(\frac{1}{e}
\{X^\mu,X^\nu, X^\sigma\}_{{\rm N}}\{X_\mu,X_\nu, X_\sigma\}_{{\rm N}}-e\Bigr), 
\label{membraneaction}
\end{equation}
\begin{equation}
\{X^\mu,X^\nu, X^\sigma\}_{{\rm N}}\equiv \sum_{a,b,c}\epsilon^{abc}
\partial_aX^\mu\partial_bX^{\nu}\partial_cX^{\sigma}, 
\label{NPbracket}
\end{equation}
giving the Dirac-Nambu-Goto form 
when the auxiliary variable $e$ is eliminated.  
Note that $\xi^a$ $(a, b, c\in (1,2,0))$ parametrize 
the 3 dimensional world volume of a single membrane, 
and space-time indices $\mu,\nu, \ldots$ run over 
11 directions of the target space-time. 
This is analogous to the 
treatments of strings where 
Poisson bracket plays a similar role \cite{schild}. 

In ref. \cite{almy} we proposed two possibilities of quantization: 
one was 
to use the ordinary square matrices and their commutators, 
and the other was more radically
 to introduce new objects, cubic matrices with three indices. 
A natural idea seemed to regularize the above action (\ref{membraneaction}) 
  directly 
by replacing the NP bracket by a finitely discretized version and 
the integral over the world volume
 by an appropriate ``Trace" operation in the 
algebra of quantized coordinates corresponding to 
classical coordinates $X^{\mu}(\xi)$. The usual light-front 
action should appear as a result of an appropriate gauge fixing of a 
higher gauge symmetry which generalizes its 
continuous counterpart, 
the area-preserving diffeomorphism 
transformations formulated a la Nambu's mechanics
\begin{equation}
\delta X^\mu=\{F,G,X^\mu\}_{{\rm N}}, 
\label{Nambutrans}
\end{equation}
with $(F(\xi), G(\xi))$ being two independent local gauge 
parameter-functions. 
At that time, we could not accomplish this 
program. One of the stumbling blocks was 
our tacit demand that the light-front time 
coordinate should also emerge automatically in the process 
of gauge fixing. This seemed to be necessary because (\ref{NPbracket}) 
involves a time derivative.  

In the present work, we reconsider the program of the covariantization 
of M(atrix) theory.\footnote{For examples of other attempts of applying 
Nambu brackets towards extended formulations  
of Matrix theory, see {\it e.g.} \cite{msato} and references therein. 
For earlier and different approaches related to our 
subject, see \cite{am} most of which discussed only the bosonic part, and 
more recent works \cite{bandos}, based on the so-called `super-embedding' method, the latter of which however 
 introduced only SO(9) matrices in contrast to one of basic requirements stressed in the present paper. 
} However, we do not pursue the above mentioned analogy with the theory of super membrane
 too far. In particular, we do {\it not} assume the above relation 
between the membrane action and Nambu bracket. Such an analogy 
does not seem to be
 essential from the viewpoint of the DLCQ interpretation with finite $N$,  
since this analogy suggests the covariance could only be 
recovered in a large $N$ limit. 
We use Nambu-type transformations only as a convenient tool to 
motivate higher gauge symmetries which would be 
necessarily required for achieving manifest covariance using 
11 dimensional matrix variables: an 
appropriate gauge-fixing of such higher gauge symmetries 
would lead us to the usual light-front theory with 9 dimensional matrix 
variables. 

With regards to the 
problem of the emergence 
of time parameter describing the causal dynamics of 
matrices, we reset our goal at a lower level. 
Namely, we introduce from the outset a single Lorentz invariant  (proper) time
parameter  $\tau$ together with an ``ein-bein" auxiliary variable $e(\tau)$, 
which transforms as 
$d\tau e(\tau)=d\tau'e'(\tau')$ under an arbitrary 
re-parametrization $\tau\rightarrow \tau'$ and generates the mass-shell condition for the 
center-of-mass variables with an 
effective mass-square operator. Thus the proper-time is essentially 
associated with the trajectory of the center-of-mass. 
From the viewpoint of relativistically covariant 
formulation of many-body systems in the 
{\it configuration-space} picture, as opposed to 
the usual second-quantized-field theory picture, we would expect that 
the proper time-parameter should be associated independently 
with each particle degree of freedom, since we 
have to impose mass-shell conditions 
separately to each particle.\footnote{For instance, we can recall 
the old many-time formalism \cite{dfp}.  It should be remembered that the usual Feynman-diagram method is a version of covariant many-body theories in configuration 
space. The Feynman parameters or Schwinger parameters play 
the role of proper times introduced for each world line 
separately. It is also to be recalled that one of the 
Virasoro constraints, $P^2+(X')^2=0$, in string theory (and the similar 
constraints in membrane theory) can be 
viewed as a counterpart of the mass-shell condition, imposed at each 
points on world sheets (or volumes).} This is possible in the usual 
relativistic quantum mechanics where we can 
separately treat particle degrees of freedom and 
field degrees of freedom which mediate interactions 
among particles, especially using Dirac's interaction representation. 
However, in matrix models such as super Yang-Mills 
quantum mechanics, such a separation is not feasible, since the 
SU($N$)  gauge symmetry associated with matrices 
requires us to treat the coordinate 
degrees and interaction degrees of freedom embedded together in each 
matrix inextricably as a single entity. In fact, in either case of 
M-theory compactifications formulated by the super Yang-Mills quantum 
mechanics, there is no trace of such mass-shell conditions set
 independently for each constituent parton. In our approach, 
the time parameters ({\it not} physical time 
components) of all the dynamical degrees of freedom are by definition 
synchronized globally to a single invariant Lorentz-invariant
 parameter of the 
center-of-mass degrees of freedom. 
Under this circumstance, we extend a higher gauge 
symmetry exhibited in our version of quantized 
Nambu bracket, and argue that it can lead to a 
mechanism for formulating many-body systems covariantly in a 
configuration-space formalism without negative metric,  replacing methods with many independent 
proper-time parameters, and hopefully 
characterizing the peculiar general-relativisitic nature of D-particles 
as partons of M-theory. 

In section 2, we first reformulate, with some slight 
extensions, our old proposal for a 
discretized Nambu bracket using matrix commutators in terms 
of ordinary square matrices to 
motivate higher gauge symmetries, and introduce a covariant 
canonical formalism to develop higher gauge transformations. 
 In section 3, we present the bosonic 
part of our action. We discuss various symmetry 
properties of the action and their implications. 
In particular, it will be demonstrated that our 
theory reduces to the usual formulation of Matrix theory in 
a light-front gauge. In section 4, we extend our theory minimally to a 
supersymmetric theory, with some details being relegated to two 
appendices. 
In section 5,  we summarize our work and conclude by mentioning various future possibilities and 
confronting problems.

\section{Canonical formalism of higher gauge symmetries}
In the present and next sections, for the purpose of elucidating the basic ideas and 
formalisms step by step in a simple setting without complications of fermionic 
degrees of freedom, we restrict ourselves to bosonic 
variables. Extension to including fermionic 
variables in a supersymmetric fashion will be 
discussed later. 

In the first part, we start from briefly recapitulating our old proposal for a discretized 
version of the Nambu bracket in the matrix form  
as a motivation toward higher gauge symmetries, and 
then in the sequel we will extend further and complete the higher gauge symmetries in 
the framework of  
a first-order canonical formalism in a relativistically covariant fashion. 

\subsection{From a discretized Nambu 3-bracket to a 
higher gauge symmetry}

Let us denote $N\times N$ hermitian matrix variables using slanted boldface symbol, 
like $\boldsymbol{X}, \boldsymbol{Y}, \boldsymbol{Z}, 
\cdots$, and introduce non-matrix variables associated with 
them and 
denoted by a special subscript M, like
$X_{\uhdb}, Y_{\uhdb}, Z_{\uhdb}, \cdots$.
All these variables are functions of the invariant time parameter $\tau$ and assumed to be scalar with respect to 
its re-parameterization. 
When we 
deal with matrix elements explicitly, we designate them by 
 $X_{ab}, \ldots $ without 
boldface symbol. 
Originally in ref. \cite{almy} we identified the $X_{\uhdb}$'s  to be 
the traces of the corresponding matrices. But that is not 
necessary, and in the present work we treat them 
as new independent dynamical degrees of freedom.\footnote{This situation itself is similar to the so-called ``Lorentzian" version of 3-algebra, 
 which is however {\it nothing} to do with our sense of the 11 dimensional Lorentzian 
symmetry of space-time, 
  It was applied in 
attempting to extend the BLG model of conformal 
field theory for M2 branes as a possible effective low 
energy description for infinitely extended multiple 
M2 branes in an SO(8)-invariant fashion. See {\it e.g.} \cite{lorentzblg} and 
references therein. Our interpretation and 
treatment are quite different from such attempts. 
In our canonical treatment 
no indefinite metric appears, except for the 
usual space-time Lorentz indices.  } 
This is the price we have to pay 
to realize a higher gauge symmetry, but we will have a 
reward too. Treating them as a pair of non-matrix and matrix 
variables, we denote like $X=(X_{{\rm M}}, \boldsymbol{X})$ 
for notational brevity.  


The discretized NP bracket, which we simply call 
3-bracket, is then defined as \footnote{This was motivated 
from Nambu's definition of a triple commutator 
$ 
\boldsymbol{X}[\boldsymbol{Y},\boldsymbol{Z}]+
\boldsymbol{Y}[\boldsymbol{Z}, \boldsymbol{X}]
+\boldsymbol{Z}[\boldsymbol{X},\boldsymbol{Y}]$ which, 
however, does {\it not} satisfy the FI.}
\begin{align}
[X, Y, Z]
\equiv (0, 
X_{\uhdb}[\boldsymbol{Y}, \boldsymbol{Z}]
+Y_{\uhdb}[\boldsymbol{Z}, \boldsymbol{X}]
+Z_{\uhdb}[\boldsymbol{X}, \boldsymbol{Y}]).
\label{almy}
\end{align}
Note that the M-component of $[X,Y,Z]$ is zero 
by definition. 
This is totally skew-symmetric and 
satisfies the so-called Fundamental Identity (FI) essentially as a 
consequence of the usual Jacobi identity, 
\begin{align}
[F,G,[X,Y,Z]]=[[F,G,X],Y,Z]+[X,[F,G,Y],Z]+[X,Y,[F,G,Z]].
\label{FI}
\end{align} 
The proof given in ref. \cite{almy}, to which we refer readers 
for further details and relevant literature related to this identity, goes through as it stands for our 
slightly extended cases too.
In particular, the absence ($[X, Y, Z]_{{\rm M}}=0$) of the M-component 
for the 3-bracket follows from 
the property that, for the matrix part of the right-hand side of (\ref{FI}), 
the contributions involving the commutator $[\boldsymbol{F}, \boldsymbol{G}]$ cancel out among themselves {\it without} performing any  trace operations for arbitrary sef of three 
elements $(X, Y, Z)$,\footnote{The reason why the proof given 
in \cite{almy} is compatible with the present extension 
is nothing more than an accidental fact that the trace of the matrix component in (\ref{almy}) also vanishes 
trivially, so that formally no contradiction arises  
even if we identify $X_{{\rm M}}$ with the trace 
of the corresponding matrix component. But the latter identification 
is not directly necessary for the validity of the proof, as explained 
in the text. } guaranteeing the absence of 
the term $[X,Y,Z]_{{\rm M}}[\boldsymbol{F}, 
\boldsymbol{G}]$.  The latter would correspond to 
the last term in the matrix part of (\ref{almy}) and, if non-vanishing,  contradict 
the vanishing of $[X,Y,Z]_{{\rm M}}$ on the left-hand side of the FI.  

If we interpret the bracket $[F,
G,X]$ for arbitrary 
variable $X$ as an infinitesimal 
gauge transformation with generators $F$ and $G$, which are local with respect to 
the proper time $\tau$, 
\begin{align}
\delta X&\equiv i[F,G,X]=(0, i[F_{\uhdb}\boldsymbol{G}-G_{\uhdb}\boldsymbol{F}, 
\boldsymbol{X}]+i[\boldsymbol{F},\boldsymbol{G}]X_{\uhdb})
\label{gaugetrans}
\end{align}
as a generalization of (\ref{Nambutrans}), 
the FI is nothing but the distribution law of gauge 
transformations for 3-bracket. 
Without losing generality, we define that the 
gauge-parameter matrix functions $\boldsymbol{F}$ 
and $\boldsymbol{G}$ are both traceless. 
An important characteristic property \cite{almy} of 
this gauge transformation is that it enables us 
to gauge away the {\it traceless} part of one of the matrix 
variables whenever its $\uhdb$ component is 
not zero, due to the second term 
in (\ref{gaugetrans}). 
On the other hand, it should be kept in mind that both 
the trace-part of the matrices and $X_{\uhdb}$ are  inert 
(${\rm Tr}(\delta \boldsymbol{X})=0=\delta X_{\uhdb}$ ) against the 
gauge transformations (\ref{gaugetrans}). 
We will later extend the 
gauge transformation slightly such that the center-of-mass 
coordinate (but still not for $X_{\uhdb}$) is also subject to 
extended gauge transformations. 

Actually, it is useful to generalize the above gauge transformation to
\begin{align}
\delta X=i\sum_r[F^r,G^r, X]=\bigl(0, \sum_r
i[F_{{\rm M}}^r\boldsymbol{G}^r-G_{{\rm M}}^r\boldsymbol{F}^r, 
\boldsymbol{X}]
+i\sum_r[\boldsymbol{F}^r, \boldsymbol{G}^r]X_{{\rm M}}\bigr)
\end{align}
by introducing an arbitrary number of independent 
gauge functions discriminated by indices $r=1, 2, \ldots$.\footnote{Such an 
extension has been mentioned already by Nambu \cite{nambu} himself in 
his attempt toward a generalized Hamiltonian mechanics.  } Since the FI (\ref{FI}) is satisfied for each $r$ separately, it is still valid after summing over them. 
This means that two traceless Hermitian matrices,  
\begin{align}
\boldsymbol{H}&\equiv \sum_r F_{{\rm M}}^r\boldsymbol{G}^r-G_{{\rm M}}^r\boldsymbol{F}^r,  \\
\boldsymbol{L}&\equiv i\sum_r[\boldsymbol{F}^r, \boldsymbol{G}^r], 
\end{align}
can be regarded as being completely independent to each other. 
In what follows, we adopt this generalized form of 
gauge transformation,
\begin{align}
\delta_{HL} X\equiv \delta_{H}X+\delta_{L}X=(0, i[\boldsymbol{H}, \boldsymbol{X}]
+\boldsymbol{L}X_{{\rm M}}), 
\end{align}
with an obvious decomposition into $\delta_{H}$ and $\delta_{L}$. 
The 3-bracket form of gauge transformation itself 
does not play any essential role for our development from this point on, 
though the 3-bracket notation will still be convenient symbolically in 
expressing action in a compact form.

For any pair of two matrices $\boldsymbol{X}, \boldsymbol{Y}$ 
with vanishing M-components $X_{\uhdb}=0=Y_{\uhdb}$, 
 the trace of their bilinear 
product
\begin{align}
\langle X, Y\rangle \equiv {\rm Tr}(\boldsymbol{X}\boldsymbol{Y}) 
\end{align}
is invariant under the gauge transformation, because the 
gauge transformation then reduces to a 
usual SU($N$) transformation $\delta_{HL} \boldsymbol{X}=i[\boldsymbol{H}, 
\boldsymbol{X}]$ and $\delta_{HL} \boldsymbol{Y}=i[\boldsymbol{H}, 
\boldsymbol{Y}]$ and hence satisfies 
a derivation property $
(\delta_{HL}\boldsymbol{X})\boldsymbol{Y}+\boldsymbol{X}
(\delta_{HL}\boldsymbol{Y})=i[\boldsymbol{H}, \boldsymbol{X}\boldsymbol{Y}]$:
\begin{align}
\delta_{HL} \langle X, Y\rangle\equiv \langle \delta_{HL} X, Y\rangle +
\langle X, \delta_{HL} Y\rangle =0.
\end{align} 
Unlike \cite{almy}, this is valid irrespectively of 
vanishing or non-vanishing trace of matrices, due to our 
treatment of $X_{{\rm M}}$'s as independent variables. 
Since the 3-brackets of an arbitrary set of matrices
 always satisfy this condition of vanishing M-component as emphasized above,  
we have a non-trivial 
gauge invariant, 
\begin{align}
\langle [X,Y,Z], [U, V, W]\rangle
\label{3brainvariant}
\end{align}
for arbitrary 
six variables $X, Y, \cdots, W$, due to the 
FI (\ref{FI}). It is to be kept in mind that for the products of 
matrices with (either and/or both) 
non-vanishing $\uhdb$-components,  
the gauge transformation does {\it not}  
satisfy the derivation property, and consequently that the 
traces of their products are not in general gauge invariant.  
This constrains systems if we require 
symmetry under our gauge transformations.  

\subsection{Coordinate-type variables}

Now we extend a 
higher gauge symmetry exhibited in the previous subsection 
within the framework of ordinary canonical formalism. 
To represent the dynamical degrees of freedom in space-time, we endow them with 
 (11 dimensional) space-time Lorentz indices 
$\mu, \nu, \sigma, \cdots$. The {\it generalized}  
coordinate vectors of D-particles are symbolized as 
$X^{\mu}=(X_{\uhdb}^{\mu}, \boldsymbol{X}^{\mu})$ 
by following the above convention.  
Their gauge transformations are 
\begin{align}
\delta_{HL} X_{{\rm M}}^{\mu}=0, \quad \delta_{HL}\boldsymbol{X}^{\mu}=
i[\boldsymbol{H}, \boldsymbol{X}^{\mu}]+\boldsymbol{L}X_{{\rm M}}^{\mu}, 
\end{align}
with $\boldsymbol{H}$ and $\boldsymbol{L}$ being 
traceless and scalar matrices. 
Thus we have a typical invariant 
$\langle [X^{\mu}, X^{\nu}, X^{\sigma}], 
[X_{\mu}, X_{\nu}, X_{\sigma}]\rangle$ involving the coordinate-type 
variables. 
The center-of-mass 
coordinate 
vector of $N$ partons is  
 $X^{\mu}_{\circ}$ which can be defined independently 
 of $N$ and designated with a special subscript $\circ$ as 
 \begin{align}
X_{\circ}^{\mu}\equiv \frac{1}{N}{\rm Tr}(\boldsymbol{X}^{\mu}), 
\quad 
\boldsymbol{X}^{\mu}=X_{\circ}^{\mu}+\hat{\boldsymbol{X}}^{\mu}, 
\quad 
{\rm Tr}(\hat{\boldsymbol{X}}^{\mu})=0
\end{align}
with $\hat{\boldsymbol{X}}^{\mu}$ being the traceless part. 
We will suppress the superscript $\, \hat{}\, $ for matrices  which are 
defined to be traceless from the beginning, unless otherwise stated. 


Since these dynamical variables in general are functions of the proper-time 
parameter $\tau$,  
we need to define covariant derivatives in order to have gauge-invariant kinetic terms. From the matrix form 
(\ref{gaugetrans}), we are led to introduce 
two kinds of {\it traceless} matrix fields as gauge fields, 
each corresponding to $\boldsymbol{H}$ and $\boldsymbol{L}$, 
which we denote by $\boldsymbol{A}$ and $\boldsymbol{B}$, respectively.  
Then, the covariant derivative is defined as
\begin{align}
\frac{D'X^{\mu}}{D\tau}&=\Bigl(\frac{dX_{\uhdb}^{\mu}}{d\tau}, 
\frac{D'\boldsymbol{X}^{\mu}}{D\tau}\Bigr),  \\
\frac{D'\boldsymbol{X}^{\mu}}{D\tau}&=
\frac{d\boldsymbol{X}^{\mu}}{d\tau}+ie[\boldsymbol{A}, 
\boldsymbol{X}^{\mu}]-e\boldsymbol{B}X^{\mu}_{\uhdb}.
\end{align}
The gauge transformations of the gauge fields are 
\begin{align}
\delta_{HL}\boldsymbol{A}&=i[\boldsymbol{H},\boldsymbol{A}]-\frac{1}{e}\frac{d}{d\tau}\boldsymbol{H}\equiv -\frac{1}{e}
\frac{D\boldsymbol{H}}{D\tau}, \\
\delta_{HL}\boldsymbol{B}&=i[\boldsymbol{H}, \boldsymbol{B}]-i[\boldsymbol{L},\boldsymbol{A}]+\frac{1}{e}\frac{d}{d\tau}\boldsymbol{L}
\equiv i[\boldsymbol{H},\boldsymbol{B}]+
\frac{1}{e}\frac{D\boldsymbol{L}}{D\tau}, 
\end{align}
resulting, in conformity with (\ref{gaugetrans}), 
\begin{align}
\delta_{HL} \Bigl(\frac{D'X^{\mu}}{D\tau}\Bigr)=
(0, \sum_r[F^r,G^r,\frac{D'X^{\mu}}{D\tau}]). 
\end{align}
Note that $\frac{D'X_{\circ}^{\mu}}{D\tau}=
\frac{dX_{\circ}^{\mu}}{d\tau}$ since $\delta_{HL}X_{\circ}^{\mu}=0$. 
The symbol $D'$ with $'$ indicates that the definition of this covariant derivative will be generalized later, taking into account further 
extensions of gauge transformations. 
It is to be kept in mind that $A_{\uhdb}$ and $B_{\uhdb}$ are zero by definition and 
also that we introduced the ein-bein $e$ in order to 
render these expressions manifestly covariant 
under re-parametrization of $\tau$, assuming that the 
gauge fields are scalar under the re-parametrization as well as 
Lorentz transformations.

It is perhaps here appropriate to pay attention to a 
possible interpretation of the mysterious 
additional vector $X_{{\rm M}}^{\mu}$. 
From the viewpoint of 11 dimensional supergravity, 
the embedding of the (type IIA) string theory built on a 
flat 10 dimensional Minkowski space-time necessitates 
specifing a background 11-dimensional metric with appropriate 
boundary conditions. Remember that the dilaton (and hence, the 
string coupling $g_s$) emerges in this process.  Consequently, it tacitly introduces a 
particular Lorentz frame in 11 dimensional Minkowski space-time. 
The vector $X_{{\rm M}}^{\mu}$ can be regarded as playing 
a similar role in our covariantized Matrix theory, and for this reason we call 
$X_{{\rm M}}^{\mu}$ and its conjugate momentum $P_{{\rm M}}^{\mu}$ to be introduced below ``M-variables": hence, with the subscript ``M". 
We assume that $X_{\uhdb}^{\mu}$ is a conserved vector, and also that 
just as the 10-dimensional background metrics and boundary conditions which are not Lorentz invariant 
 are subject to 11-dimensional 
Lorentz transformations,  
the M-variables transform as dynamical 
vector variables. Further remarks on the role 
of the M-variables will be given in section 3.

\subsection{Momentum-type variables}

In the present paper, we develop a Lorentz-covariant first-order 
formalism by introducing the 
 conjugate momenta as {\it independent} dynamical variables. 
In other words,  we 
use a Hamiltonian formalism with respect to the Lorentz-invariant 
proper time $\tau$. 
The canonical conjugates of the generalized coordinates are 
denoted by
\begin{align}
P^{\mu}=(P_{\dhd}^{\mu}, \boldsymbol{P}^{\mu}), 
\end{align}
where $P_{\dhd}^{\mu}$ and $\boldsymbol{P}^{\mu}$ 
are conjugate to $X^{\mu}_{\uhdb}$ and 
$\boldsymbol{X}^{\mu}$, respectively. 
The equal-time canonical Poisson algebra are\footnote{Our Lorentz metric is $(1,1, \cdots, 1, -1)$. }, exhibiting matrix indices explicitly, 
\begin{align}
\{X^{\mu}_{\uhdb}, P_{\dhd}^{\nu}\}_{{\rm P}}&=
\eta^{\mu\nu}, \\
\{ X^{\mu}_{ab}, P^{\nu}_{cd}\}_{{\rm P}}
&=\delta_{ad}\delta_{bc}\eta^{\mu\nu}, 
\end{align}
with all other Poisson brackets being zero 
({\it e.g.} $\{X_{ab}^{\mu}, P_{\dhd}^{\nu}\}_{{\rm P}}=0$, etc).

We demand that the canonical Poisson brackets are preserved by gauge transformations. The gauge symmetry of the 
canonical structure ensures us that we can consistently implement 
various gauge constraints when we quantize the system. 
On the basis of this requirement,  we can determine the  gauge transformations 
of canonical momenta uniquely for the 
{\it traceless} part of matrix variables,  together with the M-variables.
  The results are 
\begin{align}
\delta_{HL} \hat{\boldsymbol{P}}^{\mu}&=i[\boldsymbol{H},
\hat{\boldsymbol{P}}^{\mu}]=\delta_H\boldsymbol{P}^{\mu}, 
\label{matrixmomentumtrans}\\
\delta_{HL} P_{\dhd}^{\mu}&=
-{\rm Tr}\Bigl(
\boldsymbol{L}\boldsymbol{P}^{\mu}\Bigr)=\delta_LP_{{\rm M}}^{\mu}.
\label{auxmomentumtrans}
\end{align}
The mixing of $\boldsymbol{P}^{\mu}$ into $P^{\mu}_{\dhd}$ 
exhibited in (\ref{auxmomentumtrans}), which is 
the counterpart to the mixing of $\boldsymbol{X}^{\mu}$ and 
$X_{\uhdb}^{\mu}$ in the coordinate part, 
is necessary to guarantee the
 vanishing of $\delta_{HL} \{X_{ab}^{\mu}, P_{\dhd}^{\nu}\}_{{\rm P}}$:
\begin{align}
\delta_{HL} \{X_{ab}^{\mu}, P_{\dhd}^{\nu}\}_{{\rm P}}&=
L_{ab}\eta^{\mu\nu}-{\rm Tr}\Bigl(
\boldsymbol{L}\{X_{ab}^{\mu}, \boldsymbol{P}^{\nu}\}_{{\rm P}}\Bigr)=0.
\end{align}
It should be kept in mind that the 
laws of gauge transformation are {\it different}  
between the coordinate-type and momentum-type variables. 
In particular, the transformation law (\ref{matrixmomentumtrans}) 
ensures that the ordinary traces such as ${\rm Tr}(\boldsymbol{P}^{\mu}
\boldsymbol{P}_{\mu})$ of products of purely momentum 
variables are gauge invariant, as opposed to those involving 
the coordinate-type matrices. 

For arbitrary functions $O=O( X_{\uhdb}, \boldsymbol{X}, 
P_{\dhd}, \boldsymbol{P})$ of the generalized coordinates and momenta, the gauge transformation is expressed as a canonical 
transformation $\delta_{HL}O=\{O, {\cal C}_{HL}\}_{{\rm P}}$ in terms of an infinitesimal generator defined as 
\begin{align}
{\cal C}_{HL}\equiv {\rm Tr}
\Bigl(
\boldsymbol{P}_{\mu}\bigl(i[\boldsymbol{H}, \boldsymbol{X}^{\mu}]+\boldsymbol{L}X_{\uhdb}^{\mu}
\bigr)\Bigr),
\label{canogene}
\end{align}
making the invariance of canonical structure under 
the gauge transformations manifest. 
We note that our canonical transformations are 
explicitly proper-time dependent through 
time-dependent $\boldsymbol{H}$ and 
$\boldsymbol{L}$. In the usual 
canonical formalism, such a time-dependent 
canonical transformation 
changes the Hamiltonian 
by a shift
\begin{align}
\frac{\partial}{\partial \tau}{\cal C}_{HL}
\equiv 
{\rm Tr}
\Bigl(
\boldsymbol{P}_{\mu}\bigl(i\Bigl[\frac{d\boldsymbol{H}}{d\tau}, \boldsymbol{X}^{\mu}\Bigr]+\frac{d\boldsymbol{L}}{d\tau}X_{\uhdb}^{\mu}
\bigr)\Bigr).
\end{align}
In our generalized relativistically-invariant 
canonical formalism, this 
shift-type contribution is cancelled by the 
transformations of gauge fields. 
This is reasonable since the Hamiltonian in our system 
is zero after all, giving the {\it Hamiltonian} constraint associated with 
re-parametrization invariance with respect to $\tau$.

Being associated with these transformation laws, the 
covariant derivatives of momentum variables are 
\begin{align}
\frac{D'\boldsymbol{P}^{\mu}}{D\tau}
&\equiv \frac{d\boldsymbol{P}^{\mu}}{d\tau}+
ie[\boldsymbol{A}, \boldsymbol{P}^{\mu}], 
\label{covdev3}\\
\frac{D'P_{\dhd}^{\mu}}{D\tau}&\equiv 
\frac{dP_{\dhd}^{\mu}}{d\tau}+e{\rm Tr}(
\boldsymbol{B}\boldsymbol{P}^{\mu}),
\label{covdev4}
\end{align}
satisfying 
\begin{align}
\delta_{HL} \Bigl(\frac{D'\boldsymbol{P}^{\mu}}{D\tau}\Bigr)&=
i[
\boldsymbol{H}, \frac{D'\boldsymbol{P}^{\mu}}{D\tau}], \\
\delta_{HL} \Bigl(\frac{D'P_{\dhd}^{\mu}}{D\tau}\Bigr)&=-{\rm Tr}\Bigl(
\boldsymbol{L}\frac{D'\boldsymbol{P}^{\mu}}{D\tau}
\Bigr).
\end{align}

It is important here to notice that these canonical structure  and 
the associated covariant derivatives 
are invariant under a {\it global} (not as a local re-parametrization) scaling transformation $\tau
\rightarrow \lambda^2\tau$ of the proper time, when 
the dynamical variables are transformed as
\begin{align}
&\boldsymbol{X}^{\mu}
\rightarrow \lambda \boldsymbol{X}^{\mu}, 
\quad 
X^{\mu}_{\uhdb} \rightarrow \lambda^{-3}X_{\uhdb}^{\mu},\\
&\boldsymbol{P}^{\mu}
\rightarrow \lambda^{-1}\boldsymbol{P}^{\mu}, 
\quad 
P_{\dhd}^{\mu}\rightarrow 
\lambda^3P^{\mu}_{\dhd},\\
&
\boldsymbol{A}\rightarrow \lambda^{-2}\boldsymbol{A}, 
\quad 
\boldsymbol{B}
\rightarrow \lambda^2\boldsymbol{B}.
\end{align}
Accordingly, the gauge functions must be scaled as 
\begin{align}
\boldsymbol{H}\rightarrow \boldsymbol{H}, \quad 
\boldsymbol{L}\rightarrow \lambda^4\boldsymbol{L}.
\end{align}
Note that, by definition, the ein-bein $e$ has zero-scaling dimension,  
{\it i.e.} 
$e\rightarrow e$ and also that the canonical structure alone 
cannot fix uniquely the scaling dimensions of 
 M-variables relative to those of the matrices and $\tau$. We have chosen 
these scale dimensions such that the representative invariants 
such as $\langle [X^{\mu},X^{\nu},X^{\sigma}], [X_{\mu},X_{\nu},X_{\sigma}]\rangle$ and ${\rm Tr}(\boldsymbol{P}^{\mu}\boldsymbol{P}_{\mu})$ mentioned already 
are allowed to be main ingredients for the action. 
We also remark that this scaling symmetry is 
a disguise of the ``generalized conformal symmetry" which 
was motivated by the concept of a space-time uncertainty relation and 
 advocated in ref. \cite{jy} \footnote{The scaling transformation introduced in ref. \cite{jy} is obtained from the present definition if we redefine the proper time parameter $ed\tau =ds$ by $s=2Nx^+/P_{\circ}^+$ (see section 3) with $P_{\circ}^+=2N/R_{11}=2N/(g_s\ell_s)$ and 
then trade off the scaling $X_{{\rm M}}^{\mu}\rightarrow \lambda^{-3}
X_{{\rm M}}^{\mu}$ for $g_s\rightarrow \lambda^3g_s$ such that the transformation of $x^+$ become $x^+\rightarrow \lambda^{-1}x^+$. 
As we will see later, we can identify $\ell_{11}^{-3}=\sqrt{X_{{\rm M}}^2}=1/
(g_s\ell_s^3)$. The reader might feel here that 
in view of the signs of the scaling dimensions of $X_{{\rm M}}^{\mu}$ and 
$P_{{\rm M}}^{\mu}$ it sounds more natural to interchange the 
naming of generalized coordinate and momentum for the M-variables.  
}
in exploring gauge/gravity correspondences in the cases of 
dilatonic D-branes and scale {\it non}-invariant 
super Yang-Mills theories. It indeed 
played a useful role, for instance, 
in classifying the behavior of correlation functions in the 
context of the light-front Matrix theory in \cite{SY}. 

Corresponding to 
the invariance of canonical Poisson brackets, we 
now have a generalized one-dimensional 
Poincar\'{e} bilinear integral  
\begin{align}
\int d\tau \Bigl[
P_{\dhd\, \mu}\frac{dX_{\uhdb}^{\mu}}{d\tau}
+{\rm Tr}\Bigl(
\boldsymbol{P}_{\mu}\frac{D'\boldsymbol{X}^{\mu}}{D\tau}\Bigr)
\Bigr]=
\int d\tau \Bigl[
P_{\dhd\, \mu}\frac{dX_{\uhdb}^{\mu}}{d\tau}
+P_{\circ\, \mu}\frac{dX_{\circ}^{\mu}}{d\tau}+{\rm Tr}\Bigl(
\hat{\boldsymbol{P}}_{\mu}\frac{D'\hat{\boldsymbol{X}}^{\mu}}{D\tau}\Bigr)
\Bigr], \label{poincare1}
\end{align}
which enjoys symmetries under all the transformations introduced up to this point. 
On the right-hand side, we have separated 
the center-of-mass part, with
\begin{align}
\boldsymbol{P}^{\mu}=\frac{1}{N}P_{\circ}^{\mu}+
\hat{\boldsymbol{P}}^{\mu}, \quad 
P_{\circ}^{\mu}\equiv {\rm Tr}(\boldsymbol{P}^{\mu}).
\end{align}
Up to a total derivative this is equal to 
\begin{align}
-\int d\tau \Bigl[
\frac{D'P_{\dhd\, \mu}}{D\tau}X_{\uhdb}^{\mu}
+{\rm Tr}\Bigl(
\frac{D'\boldsymbol{P}_{\mu}}{D\tau}\boldsymbol{X}^{\mu}\Bigr)
\Bigr]
=-\int d\tau \Bigl[
\frac{D'P_{\dhd\, \mu}}{D\tau}X_{\uhdb}^{\mu}
+\frac{dP_{\circ\, \mu}}{d\tau}X_{\circ}^{\mu}
+{\rm Tr}\Bigl(
\frac{D'\hat{\boldsymbol{P}}_{\mu}}{D\tau}\hat{\boldsymbol{X}}^{\mu}\Bigr)
\Bigr].\label{poincare2}
\end{align}
Because of the above mixing, it is essential to 
treat the matrix and non-matrix components 
of generalized momenta as a single entity, as was the case of 
generalized coordinates, except for the trace components 
of the matrices which do not participate in the above 
gauge symmetry. 

We stress 
that except for the Lorentz metric $\eta^{\mu\nu}$ 
the metric appearing in the Poisson bracket, which 
upon quantization fixes the metric of Hilbert space,  
is the standard 
one. 
On the other hand, we have to take care of possible dangers of 
ordinary indefiniteness associated with 
the Minkowski nature of 11 dimensional target space.  
With respect to the center-of-mass motion, the Hamiltonian constraint arising from the variation $\delta e$ gives the mass-shell condition, which allows us to express time-like (or light-like) momentum in terms of spatial components. 
However, to deal with the time components of 
the traceless part of matrix variables, without independent 
proper times for them, we need further gauge symmetries as 
companions to $\delta_{HL}$.

\subsection{Completion of higher gauge symmetries}

One of the reasons why we need still higher gauge symmetries  
beyond $\delta_{HL}$, which already extended 
the usual SU($N$) gauge symmetry $\delta_H$, is that the 
unphysical gauge degrees of freedom of phase-space pairs 
of vector-like variables must be at least two for each (traceless) matrices in order to describe gravity, in analogy with 
string theory.\footnote{Heuristically, 
the Gauss constraints associated with 
the gauge field $\boldsymbol{B}$ and a new one
$\boldsymbol{Z}$ introduced below will play analogous (in fact much {\it stronger}) roles as 
the non-zero-mode parts of the 
Virasoro constraints $P^2+(X')^2=0$ and $P\cdot X'=0$, respectively, of string theory.  The zero-mode part of the 
former Hamiltonian constraint corresponds to our mass-shell constraint associated with 
ein-bein $e$. 
} This is necessary for reproducing the 
light-front M(atrix) theory which is described by SO(9) vector 
matrices and their super partners after an appropriate gauge-fixing. 
Possibility of such higher gauge symmetries reveals itself by noticing 
the existence of 
two natural conservation laws. We assume that 
the whole theory, being defined in the flat 11-dimensional 
Minkowski space-time, is symmetric under two rigid translations, namely, 
the usual coordinate translation 
$X_{\circ}^{\mu}\rightarrow X_{\circ}^{\mu}+c^{\mu}$ and, 
additionally, 
$P_{\dhd}^{\mu}\rightarrow P_{\dhd}^{\mu}+b^{\mu}$ 
in connection with the embedding of 10-dimensional string theory as 
emphasized already. 
As the equations of motion, we then have conservation laws for 
$P_{\circ}^{\mu}$ and $X_{\uhdb}^{\mu}$, 
\begin{align}
\frac{dP_{\circ}^{\mu}}{d\tau}=0, \quad 
\frac{dX_{\uhdb}^{\mu}}{d\tau}=0.
\label{twoconserv}
\end{align}
We can then consistently demand that $P_{\circ}^{\mu}$ is a 
time-like (or light-like as a limiting case) vector and, simultaneously,  
$X_{{\rm M}}^{\mu}$ is 
a space-like vector, and finally that they are orthogonal to each other, 
\begin{align}
P_{\circ}\cdot X_{\uhdb}=0. 
\label{orthogonality1}
\end{align}
Here and in what follows we often denote the Minkowskian scalar products  
by the  ``$\cdot$"  symbol and also use an abbreviation such as
$X_{{\rm M}}^2=X_{{\rm M}}\cdot X_{{\rm M}}$. 
Now the above orthogonality condition allows us to impose a condition 
on the matrix coordinates in a way that 
is invariant under the gauge transformation $\delta_{HL}\hat{\boldsymbol{X}}^{\mu}$, 
\begin{align}
P_{\circ}\cdot \hat{\boldsymbol{X}}=0, 
\label{orthogonality2}
\end{align}
which enables us to eliminate the time components of the 
{\it traceless} part of coordinate matrices. 

Since these two constraints are of first-class, we can 
treat them as the Gauss constraints associated with new 
gauge symmetries. 
Corresponding to (\ref{orthogonality1}) and (\ref{orthogonality2}), respectively, the local gauge transformations which preserve 
the canonical structure are given as
\begin{align}
\delta_{w} X_{\circ}^{\mu}&=w X_{\uhdb}^{\mu}, 
\quad 
\delta_{w} P_{\circ}^{\mu}=0, \quad 
\delta_{w} X_{\uhdb}^{\mu}=0, \quad \delta_{w} P_{\dhd}^{\mu}=-w P_{\circ}^{\mu},
\label{elltrans}
\end{align}
and 
\begin{align}
&\delta_Y\hat{\boldsymbol{X}}^{\mu}=0, \quad \delta_Y \hat{\boldsymbol{P}}^{\mu}=P_{\circ}^{\mu}\boldsymbol{Y}, \quad 
\delta_Y X_{\circ}^{\mu}=-{\rm Tr}(\boldsymbol{Y}
\hat{\boldsymbol{X}}^{\mu}), \quad 
\delta_Y P_{\circ}^{\mu}=0, 
\label{ytrans}
\end{align}
where $w$ and $\boldsymbol{Y}$ are 
an arbitrary function and an arbitrary {\it traceless} matrix function, respectively,  
as parameters of gauge transformations. 
It is to be noted that the other 
variables not shown here explicitly are all inert in both cases, 
and also that the conserved vectors $P_{\circ}^{\mu}$ and 
$X_{{\rm M}}^{\mu}$ are both gauge invariant. 
The expression (\ref{canogene}) of the canonical generator is now generalized to 
\begin{align}
{\cal C}_{H+L+Y+w}=wP_{\circ}\cdot X_{{\rm M}}+
{\rm Tr}\Bigl(-
(P_{\circ}\cdot\boldsymbol{X})\boldsymbol{Y}+
i\boldsymbol{P}_{\mu}[\boldsymbol{H}, \boldsymbol{X}^{\mu}]+(X_{\uhdb}\cdot \boldsymbol{P})\boldsymbol{L}
\Bigr).
\end{align}
We remark that, from the standpoint of the momentum-type variables, the combination $\delta_{HY}=\delta_H+\delta_Y$ 
can be regarded as the counterpart of $\delta_{HL}=\delta_H+\delta_L$ 
introduced previously from the standpoint of the coordinate-type variables:
in fact, $\delta_{HY}\hat{\boldsymbol{P}}^{\mu}$, if expressed in terms of 
3-bracket, is more akin to the original one introduced 
in \cite{almy}, in the sense that it uses the 
trace $P_{\circ}^{\mu}$ as the additional variable.  

The covariant derivatives are now, generalizing 
previous definitions with prime symbols,  
\begin{align}
&\frac{DX_{\circ}^{\mu}}{D\tau}= \frac{dX_{\circ}^{\mu}}{d\tau}-eB_{\circ}X_{\uhdb}^{\mu}+e
{\rm Tr}(\boldsymbol{Z}\hat{\boldsymbol{X}}^{\mu}), \\
&\frac{D\hat{\boldsymbol{X}}^{\mu}}{D\tau}=
\frac{d\hat{\boldsymbol{X}}^{\mu}}{d\tau}+ie[\boldsymbol{A}, 
\boldsymbol{X}^{\mu}]-e\boldsymbol{B}X_{{\rm M}}^{\mu}, \\
&\frac{DP_{\dhd}^{\mu}}{D\tau}=\frac{dP_{\dhd}^{\mu}}{d\tau}
+e{\rm Tr}\bigl((\boldsymbol{B}+B_{\circ})\boldsymbol{P}^{\mu}\bigr)
=\frac{dP_{\dhd}^{\mu}}{d\tau}
+e{\rm Tr}(\boldsymbol{B}\boldsymbol{P}^{\mu})
+eB_{\circ}P_{\circ}^{\mu}, \\
&\frac{D\hat{\boldsymbol{P}}^{\mu}}{D\tau}=
\frac{d\hat{\boldsymbol{P}}^{\mu}}{d\tau}+ie
[\boldsymbol{A}, \boldsymbol{P}^{\mu}]
-e\boldsymbol{Z}P_{\circ}^{\mu}, 
\end{align}
transforming as 
\begin{align}
&(\delta_{HL}+\delta_{w}+\delta_Y)\Bigl(\frac{DX_{\circ}^{\mu}}{D\tau}\Bigr)
=
L\frac{dX_{\uhdb}^{\mu}}{d\tau}
-{\rm Tr}\Bigl(
\boldsymbol{Y}\frac{D\hat{\boldsymbol{X}}^{\mu}}{D\tau}
\Bigr), \\
&(\delta_{HL}+\delta_{w}+\delta_Y)
\Bigl(\frac{D\hat{\boldsymbol{X}}^{\mu}}{D\tau}\Bigr)=
i[\boldsymbol{H}, \frac{D\hat{\boldsymbol{X}}^{\mu}}{D\tau}]
+\boldsymbol{L}\frac{dX_{{\rm M}}^{\mu}}{d\tau}, 
\\&(\delta_{HL}+\delta_{w}+\delta_Y)
\Bigl(\frac{DP_{{\rm M}}^{\mu}}{D\tau}\Bigr)=-{\rm Tr}
\Bigl(\boldsymbol{L}\frac{D\boldsymbol{P}^{\mu}}{D\tau}\Bigr)
-L\frac{dP_{\circ}^{\mu}}{d\tau}, 
\\
&(\delta_{HL}+\delta_{w}+\delta_Y)\Bigl(
\frac{D\hat{\boldsymbol{P}}^{\mu}}{D\tau}\Bigr)
=i[\boldsymbol{H}, 
\frac{D\hat{\boldsymbol{P}}^{\mu}}{D\tau}]+\boldsymbol{Y}\frac{
dP_{\circ}^{\mu}}{d\tau}.
\end{align}
We introduced new gauge fields $B_{\circ}$ and $\boldsymbol{Z}$ 
whose transformation laws are 
\begin{align}
&\delta_{HL}B_{\circ}={\rm Tr}(\boldsymbol{L}\boldsymbol{Z}), \\
&\delta_{HL}\boldsymbol{Z}=i[\boldsymbol{H}, \boldsymbol{Z}], \\
&\delta_{w} B_{\circ} =\frac{1}{e}\frac{dw}{d\tau}, \quad 
\delta_{w} \boldsymbol{Z}=0, \\
& \delta_Y B_{\circ}=-{\rm Tr}(\boldsymbol{Y}\boldsymbol{B}),  \\
&\delta_Y \boldsymbol{Z}=\frac{1}{e}\frac{d\boldsymbol{Y}}{d\tau}
+i[\boldsymbol{A}, \boldsymbol{Y}]\equiv \frac{1}{e}
\frac{D\boldsymbol{Y}}{D\tau}, 
\end{align}
and scalings are 
\begin{align}
B_{\circ}\rightarrow \lambda^2 B_{\circ}, \quad 
\boldsymbol{Z}\rightarrow \lambda^{-2}\boldsymbol{Z}. 
\end{align}
Like other matrix gauge fields, the matrix gauge field $\boldsymbol{Z}$ is traceless by definition. 
It is also to be kept in mind that both the conserved 
vectors $P_{\circ}^{\mu}$ and $X_{{\rm M}}^{\mu}$ 
are completely inert under all of gauge transformations. 

The schematic structure of higher gauge symmetries is summarized in 
Fig. 1.  The non-dynamical matrix gauge 
fields are defined to be traceless and hence matrix-type 
Gauss constraints are also traceless, the gauge 
structure of our model is essentially SU($N$) rather than U($N$), 
though the gauge field $B_{\circ}$ behaves partially as the trace component 
associated with the traceless matrix gauge field $\boldsymbol{B}$.
On the other hand, for 
dynamical coordinate and momentum variables, the U($1$) trace parts (or the center-of-mass parts) 
also play indispensable roles. However, as Fig. 1 suggests, the separate treatment of them is essential for the higher symmetries,  
especially $\delta_Y$,  
in realizing 11 dimensional covariance. 
The importance of such a separation will later become more evident in 
the treatment of the 
fermionic part and supersymmetries as we shall discuss in section 4.

Provided that derivative terms in the action appear 
only through the first-order generalized Poincar\'{e} integral 
\begin{align}
\int d\tau \Bigl[
P_{\dhd\, \mu}\frac{dX_{\uhdb}^{\mu}}{d\tau}
+{\rm Tr}\Bigl(
\boldsymbol{P}_{\mu}\frac{D\boldsymbol{X}^{\mu}}{D\tau}\Bigr)
\Bigr]&=
\int d\tau \Bigl[
P_{\dhd\, \mu}\frac{dX_{\uhdb}^{\mu}}{d\tau}
+P_{\circ\, \mu}\frac{DX_{\circ}^{\mu}}{D\tau}+{\rm Tr}\Bigl(
\hat{\boldsymbol{P}}_{\mu}\frac{D\hat{\boldsymbol{X}}^{\mu}}{D\tau}\Bigr)
\Bigr]\nonumber \\
&=-\int d\tau \Bigl[
\frac{DP_{\dhd\, \mu}}{D\tau}X_{\uhdb}^{\mu}
+\frac{dP_{\circ\, \mu}}{d\tau}X_{\circ}^{\mu}
+{\rm Tr}\Bigl(
\frac{D\hat{\boldsymbol{P}}_{\mu}}{D\tau}\hat{\boldsymbol{X}}^{\mu}\Bigr)
\Bigr], 
\label{poincare2}
\end{align}
which is, with generalized covariant derivatives,  now invariant under the whole set of gauge transformations, 
the Gauss contraints  are precisely (\ref{orthogonality1}) and 
(\ref{orthogonality2}), corresponding to 
the gauge fields $B_{\circ}$ and $\boldsymbol{Z}$, respectively, 
together with those associated with $\boldsymbol{B}$ and 
$\boldsymbol{A}$.

Corresponding to the manifest Lorentz covariance of the canonical 
structure, the standard form of Lorentz generators
\begin{align}
{\cal M}^{\mu\nu}\equiv 
X_{{\rm M}}^{\mu}P_{{\rm M}}^{\nu}-X_{{\rm M}}^{\nu}P_{{\rm M}}^{\mu}
+{\rm Tr}(\boldsymbol{X}^{\mu}\boldsymbol{P}^{\nu}-
\boldsymbol{X}^{\nu}\boldsymbol{P}^{\mu})
\label{lorentzgene}
\end{align}
are gauge invariant $\{{\cal M}^{\mu\nu}, {\cal C}_
{HL+w +Y}\}_{{\rm P}}=0$ and satisfy the Lorentz algebra with respect to 
the Poisson bracket. 

\begin{figure}[htbp]
\begin{minipage}[t]{0.45\textwidth}
\begin{center}
\includegraphics[width=0.95\textwidth,clip]
{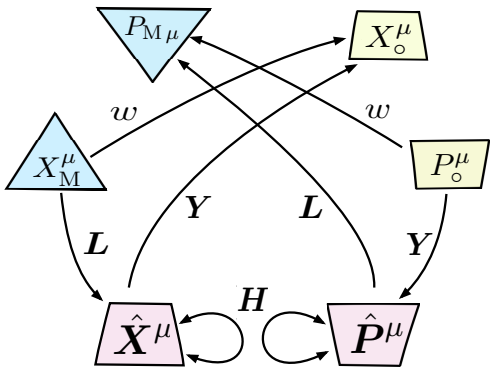}
\end{center}
\end{minipage}
\hspace{0.5cm}
\begin{minipage}[t]{0.52\textwidth}
\vspace{-5cm}
\begin{center}
\caption{{\small Schematic structure of the higher gauge symmetries: The different shapes of the objects indicate different scaling dimensions of canonical variables. The directions of arrows indicate how the variables are  mixed into others (or into themselves) by gauge transformations. 
The row in the middle represents conserved vectors, while the top row
represents the corresponding cyclic (passive) variables. 
Although superficially the transformations are acting symmetrically 
between the left and right sides of this diagram, their roles are different.  }}
\end{center}
\end{minipage}
\end{figure}

\section{Bosonic action}
We now have tools at our disposal to construct the 
action integral.  For simplicity, we still 
concentrate to the bosonic part in this section. Our basic requirement is 
that the action should have symmetries, apart from the requirement of full  SO(10,1) Lorentz-Poincar\'{e}  
invariance, under all transformations, namely, local 
$\tau$-reparametrizations, gauge transformations, as well as the 
global scale transformations and translations, which leave the canonical 
structure invariant. 
Up to total derivatives, 
unique possibility for the first-order (with respect to 
derivative) term is
the Poincar\'{e} integral (\ref{poincare2}). 
As the simplest possible potential term satisfying these requirements,  
we choose 
using (\ref{3brainvariant}), 
\begin{align}
&\frac{1}{12}\int d\tau \,e\, \langle [X^{\mu},X^{\nu},X^{\sigma}]
[X_{\mu},X_{\nu},X_{\sigma}]\rangle \nonumber \\
=&\frac{1}{4}\int d\tau \, e\, {\rm Tr}
\Big(X_{\uhdb}^2
[\boldsymbol{X}^{\nu},\boldsymbol{X}^{\sigma}][\boldsymbol{X}_{\nu},\boldsymbol{X}_{\sigma}]
-2[X_{\uhdb}\cdot \boldsymbol{X}, \boldsymbol{X}^{\nu}]
[X_{\uhdb}\cdot\boldsymbol{X}, \boldsymbol{X}_{\nu}]
\Bigr). 
\label{potential}
\end{align}
It is to be noted that the numerical 
proportional constant in front of the potential is 
arbitrary, since we can always absorb it by making a global rescaling 
$(X_{\uhdb}^{\mu}, P_{\dhd}^{\mu}) \rightarrow (\rho X^{\mu}_{\uhdb}, 
\rho^{-1} P_{\dhd}^{\mu})$, $
(B_{\circ}, \boldsymbol{B})\rightarrow \rho^{-1} (B_{\circ}, \boldsymbol{B})$ which keeps the 
the first-order term intact.

In order to have non-trivial dynamics, we need 
at least quadratic kinetic terms, typically as 
\[
-\int d\tau \, \frac{e}{2}{\rm Tr} (\boldsymbol{P}\cdot \boldsymbol{P}), 
\]
which however apparently violates gauge symmetry under (\ref{ytrans}). 
The symmetry can be recovered by the following procedure, 
which is analogous to a well known situation in the covariant 
field theory of a massive vector field.\footnote{
It may be instructive here to formulate a massive Abelian vector field in  the first-order formalism (in four dimensions) with action
\[
\int d^4x \bigl(-
\partial_{\mu}A_{\nu}F^{\mu\nu} +
\frac{1}{4}F_{\mu\nu}F^{\mu\nu}-\frac{m^2}{2}A_{\mu}A^{\mu}
\bigr).
\]
Note that we introduce an antisymmetric-tensor 
field $F_{\mu\nu}=-F_{\nu\mu}$ as an independent variable. 
The first term as an analogue to our Poincar\'{e} integral 
is invariant under two independent
 gauge transformations $\delta A_{\mu}=
\partial_{\mu}\lambda$ and $\delta F^{\mu\nu}=
\frac{1}{2}\epsilon^{\mu\nu\alpha\beta}(
\partial_{\alpha}\Lambda_{\beta}-\partial_{\beta}\Lambda_{\alpha})$ up to 
total derivative, while the 
2nd and 3rd quadratic terms are not invariant, analogously to 
 ${\rm Tr}(\boldsymbol{P}^2)$. The equations of motion 
reduce to $(\partial^2-m^2)A_{\mu}=0$ and $ \partial_{\mu}A^{\mu}=0$, 
the latter of which eliminates the negative norm. No inconsistency 
arises here. The quadratic terms 
act partially as gauge-fixing terms for the gauge symmetry 
of the first term precisely as in the system we are pursuing. As is well known, it is possible to recover the 
gauge symmetry by introducing further unphysical 
degrees of freedom, the so-called Stueckelberg field (or the `gauge part' of a Higgs field) which corresponds to our $\boldsymbol{K}$. }  
Namely, we introduce an auxiliary traceless matrix field 
$\boldsymbol{K}$ transforming simply as 
\begin{align}
\delta_{Y}\boldsymbol{K}=\boldsymbol{Y}. 
\end{align}
Then, by 
replacing $\boldsymbol{P}^{\mu}$ as $\boldsymbol{P}^{\mu}\rightarrow 
\boldsymbol{P}^{\mu}-P_{\circ}^{\mu}\boldsymbol{K}$, 
we have an invariant quadratic kinetic term, 
\begin{align}
-\int d\tau\, \frac{e}{2}{\rm Tr}(\boldsymbol{P}-P_{\circ}\boldsymbol{K})^2
=
-\int d\tau\, \frac{e}{2}\Bigl(\frac{1}{N}P_{\circ}^2 +{\rm Tr} (\hat{\boldsymbol{P}}-P_{\circ}\boldsymbol{K})^2
\Bigr). 
\end{align}
The standard kinetic term without $\boldsymbol{K}$ is obtained 
by adopting $\boldsymbol{K}=0$ as the gauge condition.  
Since the equation of `motion' (rather, another Gauss constraint) 
for $\boldsymbol{K}$ is 
\begin{align}
P_{\circ}\cdot (\hat{\boldsymbol{P}}-P_{\circ}\boldsymbol{K})=0, 
\label{keqmotion}
\end{align}
this gauge choice is actually equivalent to the following choice of 
gauge condition
\begin{align}
P_{\circ}\cdot \hat{\boldsymbol{P}}=0,  
\end{align} 
which renders the Gauss constraint (\ref{orthogonality2}) into 
a second-class constraint.

Putting together all the ingredients, the final form 
of bosonic action is 
\begin{align}
A_{{\rm boson}}&=\int d\tau 
\Bigl[
P_{\circ}\cdot \frac{DX_{\circ}}{D\tau}
+P_{\dhd}\cdot\frac{dX_{\uhdb}}{d\tau}
+{\rm Tr}\Bigl(
\hat{\boldsymbol{P}}\cdot\frac{D\hat{\boldsymbol{X}}}{D\tau}\Bigr)
\nonumber \\
&-\frac{e}{2N}P_{\circ}^2
-\frac{e}{2}\, {\rm Tr}(\hat{\boldsymbol{P}}-P_{\circ}\boldsymbol{K})^2 
+ \frac{e}{12} \bigl< [X^{\mu},X^{\nu},X^{\sigma}]
[X_{\mu},X_{\nu},X_{\sigma}]\bigr>
\Bigr].
\label{bosonaction}
\end{align}
Clearly, this is the simplest possible non-trivial form 
of the action. 
The variation of the ein-bein $e$ gives the 
mass-shell constraint for the center-of-mass momentum 
\begin{align}
 & \qquad\qquad P_{\circ}^2+{\cal M}^2_{{\rm boson}}\simeq0, 
\label{massshell}
\end{align}
with the effective invariant mass-square ${\cal M}^2_{{\rm boson}}$ being given by
\begin{align}
{\cal M}^2_{{\rm boson}}=N {\rm Tr}(\hat{\boldsymbol{P}}-P_{\circ}\boldsymbol{K})^2
- \frac{N}{6} \bigl< [X^{\mu},X^{\nu},X^{\sigma}]
[X_{\mu},X_{\nu},X_{\sigma}]\bigr>, 
\end{align}
which involves only the traceless matrices and 
is positive semi-definite {\it on-shell} with 
$\hat{\boldsymbol{P}}^{\mu}-P_{\circ}^{\mu}\boldsymbol{K}
=\frac{1}{e}\frac{D\hat{\boldsymbol{X}}^{\mu}}{D\tau}$ under the Gauss constraints, since the time component 
of the traceless matrices are eliminated by these constraints:
by the symbol $\simeq$ in (\ref{massshell}), 
we indicate that the equality is valid in conjunction with the 
Gauss-law constraints, 
\begin{align}
[\boldsymbol{P}_{\mu}, \boldsymbol{X}^{\mu}]=0, 
\label{bosongauss1}\\
\hat{\boldsymbol{P}}\cdot X_{\uhdb}=0, 
\label{bosongauss2}
\end{align}
associated with the gauge fields $\boldsymbol{A}$ and 
$\boldsymbol{B}$, respectively, together with (\ref{orthogonality1}) and 
(\ref{orthogonality2}). It should be kept in mind that ultimately,  
after taking into account fermionic contribution to be discussed in the next section, 
we are interested in states for which the 
effective mass-square is of order one in the large $N$ limit. 

In order to demonstrate that the above bosonic action has 
desirable properties as a covariantized version of Matrix theory, 
we now check some expected features. 

\vspace{0.2cm}
\noindent
(1) {\it Consistency of the Gauss constraints with the 
equations of motion}

As a first exercise, let us see briefly how the Gauss constraints (\ref{bosongauss1}) and (\ref{bosongauss2}) are consistent with the equations of motion,
\begin{align}
\frac{d\hat{\boldsymbol{P}_{\mu}}}{d\tau}
+ie[\boldsymbol{A}, 
\boldsymbol{P}^{\mu}]=eP_{\circ\, \mu}\boldsymbol{Z}-\frac{e}{2N}\frac{\partial}{\partial \hat{\boldsymbol{X}}^{\mu}}{\cal M}^2_{{\rm boson}}.
\label{eqmotion1}
\end{align}
The $\delta_{HL}$-gauge invariance of the potential is equivalent with the 
following identities. 
\begin{align}
&X_{\uhdb\, \mu}\frac{\partial}{\partial \boldsymbol{X}^{\mu}}
{\cal M}^2_{{\rm boson}}=0, \\
&[\boldsymbol{X}_{\mu}, \frac{\partial}{\partial \boldsymbol{X}^{\mu}}
{\cal M}^2_{{\rm boson}}]=0
. 
\end{align}
Then, by taking a contraction with $X_{\uhdb}^{\mu}$ and using 
(\ref{orthogonality1}) with the conservation of $X_{{\rm M}}^{\mu}$,  
(\ref{eqmotion1}) leads to 
\begin{align}
\frac{d}{d\tau}(X_{{\rm M}}\cdot \hat{\boldsymbol{P}})
+ie[\boldsymbol{A}, X_{{\rm M}}\cdot \hat{\boldsymbol{P}}]=0.
\end{align}
On the other hand, by 
taking a commutator 
with 
$\boldsymbol{X}^{\mu}$ and using the first-order equations 
of motion for it
\begin{align}
\hat{\boldsymbol{P}}^{\mu}-P_{\circ}^{\mu}\boldsymbol{K}=\frac{1}{e}
\frac{d\hat{\boldsymbol{X}}^{\mu}}{d\tau}+i[\boldsymbol{A}, 
\boldsymbol{X}^{\mu}]-X_{{\rm M}}^{\mu}\boldsymbol{B}
\label{1stXeq}
\end{align}
together with (\ref{orthogonality2}) and (\ref{keqmotion}), we can 
derive 
\begin{align}
\frac{1}{e}\frac{d}{d\tau}\bigl([\boldsymbol{X}^{\mu}, \boldsymbol{P}_{\mu}]\bigr)
=i [\boldsymbol{A}, [\boldsymbol{X}^{\mu}, \boldsymbol{P}_{\mu}]]
+[\boldsymbol{B}, X_{{\rm M}}\cdot \boldsymbol{P}], 
\end{align}
ensuring the consistency of the 
Gauss constraints (\ref{bosongauss1}) and (\ref{bosongauss2}). 
The consistency of (\ref{orthogonality1}) and 
(\ref{orthogonality2}) with the equations of motion can also be 
easily checked: the conservation of $X_{{\rm M}}$ and $P_{\circ}^{\mu}$ 
ensures the time independence of (\ref{orthogonality1}), 
while contracting $P_{\circ\, \mu}$ with (\ref{1stXeq}) 
gives 
\begin{align}
\frac{d}{d\tau}(P_{\circ}\cdot \hat{\boldsymbol{X}}) 
+ie[\boldsymbol{A}, P_{\circ}\cdot \boldsymbol{X}]-e
P_{\circ}\cdot X_{{\rm M}}\hat{\boldsymbol{B}}=0.
\end{align}

One comment relevant here is that 
the dynamical role of the M-momentum $P_{\dhd}^{\mu}$ is to lead 
the conservation of $X_{\uhdb}^{\mu}$, and that it does not 
participate in the dynamics of this system actively, 
since there is no kinetic term for it. 
Its behavior is determined by the 
equation of motion in terms of the other variables in a completely 
passive manner as
\begin{align}
\frac{DP_{{\rm M}}^{\mu}}{D\tau}=-\frac{\partial}{\partial X_{\uhdb\, \mu}}{\cal V},
\label{pdhdeq}
\end{align}
where we denoted the potential term in the action by $-{\cal V}$. 
Note that the center-of-mass coordinate 
$X_{\circ}^{\mu}$ is also of passive nature, similarly, leading to the
conservation of the center-of-mass momentum, and that 
its time derivative is expressed entirely in terms of the other variables. 
In other words, both these variables are  
``cyclic" variables using the terminology of analytical mechanics.

\vspace{0.2cm}
\noindent
(2) {\it Light-front and time-like gauge fixings}

As a next check, let us demonstrate that this system 
reduces to the bosonic part of 
 light-front Matrix theory after an 
appropriate gauge fixing together with the condition 
of compactification. Without losing generality, we first 
choose a two-dimensional (Minkowskian) plane spanned by two conserved vectors 
$P_{\circ}^{\mu}$ and $X_{\uhdb}^{\mu}$ and introduce the light-front coordinates ($P^{\pm}_{\circ}\equiv 
P^{10}_{\circ}\pm P^0_{\circ}$ and 
$X_{\uhdb}^{\pm}\equiv X^{10}_{\uhdb}\pm X^0_{\uhdb}$) foliating this plane. For convenience, we call this plane ``M-plane". Note that due to the space-like nature of $X_{\uhdb}^{\mu}$ together with the constraint (\ref{orthogonality1}), 
both of its light-front components $X_{\uhdb}^{\pm}$ are non-vanishing, 
 while by definition two conserved vectors 
$P_{\circ}^{\mu}$ and $X_{\uhdb}^{\mu}$ have 
no transverse components orthogonal to the M-plane. We can then choose the gauge using the $\delta_{L}$-gauge symmetry 
such that 
\begin{align}
\hat{\boldsymbol{X}}^+=0.
\label{lfgauge}
\end{align}
The remaining 
light-like component $\hat{\boldsymbol{X}}^-$ is in the second term of the 
potential term 
\begin{align}
\frac{1}{8}\int d\tau \, e\, {\rm Tr}
([X_{\uhdb}^+\boldsymbol{X}^-, \boldsymbol{X}^i]
[X_{\uhdb}^+\boldsymbol{X}^-, \boldsymbol{X}^i]), 
\label{xminus}
\end{align}
with $i$ running only over the SO(9) directions which are 
transverse to the M-plane. This is eliminated 
by the $\delta_{Y}$-Gauss constraint
\begin{align}
0=P_{\circ}^+\hat{\boldsymbol{X}}^-+P_{\circ}^-\hat{\boldsymbol{X}}^+=
P_{\circ}^+\hat{\boldsymbol{X}}^-\quad \Rightarrow \quad \hat{\boldsymbol{X}}^-=0, 
\end{align}
under the condition $P_{\circ}^+\ne 0$. We stress that 
without this particular constraint we cannot 
derive the potential term coinciding with the light-front Matrix 
theory. 
As for the momentum variables, we can use  
the $\boldsymbol{B}$-gauge Gauss constraint 
\begin{equation}
0=X_{\uhdb}\cdot \hat{\boldsymbol{P}}=
X^-_{\uhdb}\hat{\boldsymbol{P}}^++X_{\uhdb}^+\hat{\boldsymbol{P}}^-
\quad \Rightarrow \quad \boldsymbol{B}X_{\uhdb}^2=0 \quad 
\Rightarrow \quad \boldsymbol{B}=0, 
\end{equation}
with the assumption $X_{\uhdb}^2>0$, using the 
first-order equations of motion after choosing the 
gauge condition $\boldsymbol{K}=0$ with 
respect to the $\delta_w$-gauge symmetry, 
\begin{align}
\hat{\boldsymbol{P}}^{\pm}=\frac{1}{e}\frac{d\hat{\boldsymbol{X}}^{\pm}}{d\tau}
+i[\boldsymbol{A}, \hat{\boldsymbol{X}}^{\pm}]
-\boldsymbol{B}X_{\uhdb}^{\pm}. 
\end{align}
 The result in the end is simply 
\begin{align}
\hat{\boldsymbol{P}}^{\pm}=0, 
\end{align}
which also implies $\boldsymbol{Z}=0$ as a consequence of 
(\ref{eqmotion1}). 
Note also that the $\boldsymbol{A}$-gauge Gauss constraint takes the form
\begin{align}
[\boldsymbol{X}_i, \boldsymbol{P}_i]=0.
\end{align}

Now all light-like 
components of the traceless matrix variables are 
completely eliminated. The 
effective mass square in the light-front gauge takes the form
\begin{align}
{\cal M}^2_{{\rm bosonlf}}=
N {\rm Tr}\Bigl(\hat{\boldsymbol{P}}_i\cdot\hat{\boldsymbol{P}} _i
- \frac{1}{2}
X_{\uhdb}^2
[\boldsymbol{X}_i,\boldsymbol{X}_j][\boldsymbol{X}_i,\boldsymbol{X}_j]
\Bigr).
\end{align}
From this result, it follows that the conserved Lorentz invariant 
$X_{\uhdb}^2$ gives the 11 dimensional gravitational length as\footnote{It should be kept in mind that at this point there is no independent meaning  in separating string coupling $g_s$, which 
acquires its independent role only after imposing the condition 
of compactification. }
\begin{align}
X_{\uhdb}^2=\frac{1}{\ell_{11}^6}.
\label{11dlength}
\end{align}

The equations of motion for the center-of-mass variables and 
for $X_{\uhdb}^{\mu}$
are, using $\hat{\boldsymbol{X}}^{\pm}=0$ and setting $ds=ed\tau$, 
\begin{align}
&P_{\circ}^{\pm}=N\Bigl(\frac{dX_{\circ}^{\pm}}{ds}-B_{\circ}X_{\uhdb}^{\pm}\Bigl), 
\quad 
\frac{dP^{\pm}_{\circ}}{ds}=0, \quad 
\frac{dX_{\uhdb}^{\pm}}{ds}=0.
\end{align} 
With respect to the $\delta_{w}$-gauge symmetry, we can choose a 
gauge $B_{\circ}=0$. Then, 
\begin{align}
P_{\circ}^{\pm}=N\frac{dX_{\circ}^{\pm}}{ds} ,
\end{align}
and we can identify the re-parametrization 
invariant time parameter $s$ with the 
center-of-mass light-front time coordinate as
\begin{align}
X^+_{\circ}=\frac{P_{\circ}^+}{N}s.
\end{align}

The effective action for the remaining transverse variables is 
obtained by substituting the solutions of  constraints resulting from the 
mass-shell condition
\begin{align}
P_{\circ}^-=-\frac{{\cal M}^2_{{\rm lf}}}{P_{\circ}^+}
\end{align}
into the original action. Then, neglecting a total derivative, we obtain
\begin{align}
A_{{\rm lf\, boson}}&=\int ds \Bigl[
{\rm Tr}\Bigl(
\hat{\boldsymbol{P}}_i\frac{D\hat{\boldsymbol{X}_i}}{Ds}
\Bigr)+\frac{1}{2}P_{\circ}^-\frac{dX^+_{\circ}}{ds}
\Bigl]=\int ds \Bigl[
{\rm Tr}\Bigl(
\hat{\boldsymbol{P}}_i\frac{D\hat{\boldsymbol{X}_i}}{Ds}
\Bigr)-\frac{1}{2N}{\cal M}_{{\rm bosonlf}}^2
\Bigl]\\
&\Rightarrow 
\int ds 
{\rm Tr}\Bigl[\frac{1}{2}
\frac{D\hat{\boldsymbol{X}_i}}{Ds}\frac{D\hat{\boldsymbol{X}_i}}{Ds}+\frac{1}{4}
X_{\uhdb}^2
[\boldsymbol{X}_i,\boldsymbol{X}_j][\boldsymbol{X}_i,\boldsymbol{X}_j]\Bigl]\nonumber \\
&=\int dx^+\frac{1}{2R}{\rm Tr}\Bigl(\frac{D\hat{\boldsymbol{X}}^i}{Dx^+}\frac{D\hat{\boldsymbol{X}_i} }{Dx^+}+ \frac{R^2}{2\ell_{11}^6}[\boldsymbol{X}_i, \boldsymbol{X}_j][\boldsymbol{X}_i, \boldsymbol{X}_j]\Bigr),
\label{lfeffectiveaction}
\end{align}
where in the second line we shifted from our first-order form to the second-order 
formalism by integrating out the transverse momenta $\hat{\boldsymbol{P}}_i$, and 
in the third line, we  have 
rescaled the time coordinate by $s=2Nx^+/P_{\circ}^+$ 
($X_{\circ}^+=2x^+$) 
with the constant 
light-front momentum $P_{\circ}^+$ discretized with the 
DLCQ compactication by introducing a continuous parameter $R$ which 
can be changed arbitrarily by boost, 
\begin{align}
P_{\circ}^+=\frac{2N}{R}.
\label{lfcomp}
\end{align}
This condition expresses our premise that 
constituent partons all have the same 
basic unit $1/R$ of compactified momentum.\footnote{
As stressed in the Introduction, that $N$ as the number 
of constituent D-particles is a conserved and 
Lorentz-invariant quantum number is a fundamental 
assumption of our construction. Even though $N$ itself is 
gauge invariant by definition, its relation 
with momentum  and compactification radius depends on the choice of gauge and/or Lorentz frame. }
 Note also that it 
amounts to requiring that the relation between 
the light-front time $X_{\circ}^+$ and the 
invariant proper time $s$ is independent of $N$. Because of 
a global synchronization of the proper-time parameter as stressed 
in section 1, 
this is as it should be since the same relation 
between the target time  and the 
proper time should hold for susbsystems when the 
system is regarded as a composite of many subsystems with 
smaller $N_i$'s such that $N=\sum_iN_i$. 
The gauge 
field $\boldsymbol{A}$ is also rescaled, 
$\boldsymbol{A}\rightarrow \frac{P_{\circ}^+}{2N}\boldsymbol{A}=\boldsymbol{A}/R$, 
and the covariant derivative is now without $\boldsymbol{B}$-gauge field since $X_{\uhdb}^i=0$ as 
\begin{align}
\frac{D\hat{\boldsymbol{X}}^i}{Dx^+}=
\frac{d\hat{\boldsymbol{X}}^i}{dx^+}+i[\boldsymbol{A}, \boldsymbol{X}^i].
\end{align}
It is to be noted, as discussed in section 1, that if we 
set $R\Rightarrow R_{11}=g_s\ell_s$, this form (\ref{lfeffectiveaction}) 
is identical with
 the low-energy effective action for D-particles 
in the weak-coupling limit $g_s\rightarrow 0$, giving an 
infinite momentum frame with fixed $N$ from a viewpoint of 
11 dimensions as discussed in section 1. 

Let us also briefly consider the case of a spatial compactification. 
We use the same frame for the two-dimensional M-plane spanned by 
$P_{\circ}^{\mu}$ and $X_{\uhdb}^{\mu}$, but we foliate 
it in terms of the ordinary time coordinate $X_{\circ}^0$ and 
choose the time-like gauge
\begin{align}
\hat{\boldsymbol{X}}^0=0, 
\end{align}
which is possible since $X^0_{\uhdb}\ne 0$ under the 
requirements $P_{\circ}^{10}> 0, P_{\circ}^0>0$ due to the 
$\delta_{w}$-Gauss constraint (\ref{orthogonality1}). 
Then, the constraint (\ref{orthogonality2}) together with 
the $\boldsymbol{B}$-and-$\boldsymbol{Z}$-Gauss constraints leads 
to 
\begin{align}
\hat{\boldsymbol{X}}^{10}=0, 
\end{align}
along with the corresponding momentum-space counterparts. 
Thus, as for the longitudinal component, we have the same results as the 
light-front case. 
Only difference is that the condition of compactification is,  
instead of (\ref{lfcomp}), 
\begin{align}
P^{10}_{\circ}=\frac{N}{R_{11}},
\label{spatcomp}
\end{align}
and therefore the mass-shell constraint 
for the center-of-mass momentum is solved as
\begin{align}
P_{\circ}^{0}
=
\sqrt{(P_{\circ}^{10})^2+N{\rm Tr}\Bigl(\hat{\boldsymbol{P}}_i\cdot\hat{\boldsymbol{P}} _i
- \frac{1}{2}
X_{\uhdb}^2
[\boldsymbol{X}_i,\boldsymbol{X}_j][\boldsymbol{X}_i,\boldsymbol{X}_j]
\Bigr)},
\end{align}
which 
leads to the effective action
\begin{align}
A_{{\rm spat\, boson}}=
\int dt \Bigl[{\rm Tr}\Bigl(
\hat{\boldsymbol{P}}_i\frac{D\hat{\boldsymbol{X}}_i}{Dt}\Bigr)
-P_{\circ}^0
\Bigr],
\end{align}
where we changed the parametrization by $t=X_{\circ}^0=
\frac{P_{\circ}^{10}}{N}s=s/R_{11}$ and made a rescaling of the 
gauge field $\boldsymbol{A}$ correspondingly. 
On shifting to a first-order formalism by solving 
the momenta $\hat{\boldsymbol{P}}_i$ in terms of 
the coordinate variables, we arrive at a 
Born-Infeld-like action
\begin{align}
A_{{\rm spat\, boson}}=-
\int dt\,  {\cal M}_{{\rm spat}}\sqrt{N}\Bigl[
1-\frac{1}{N}{\rm Tr}\Bigl(\frac{D\hat{\boldsymbol{X}}_i}{Dt}
\frac{D\hat{\boldsymbol{X}}_i}{Dt}\Bigr)\Bigr]^{1/2}
\label{timelikeaction}
\end{align}
with 
\begin{align}
{\cal M}_{{\rm spat}}\equiv \Bigl[\frac{N}{R_{11}^2}
-\frac{1}{2\ell_{11}^6}{\rm Tr}\bigl([\boldsymbol{X}_i,\boldsymbol{X}_j]
[\boldsymbol{X}_i,\boldsymbol{X}_j]\bigr)\Bigr]^{1/2}, 
\label{tmbosonpotential}
\end{align}
which, in the limit of large $N$, can be approximated by
\begin{align}
\int dt\, \frac{N}{R_{11}}\Bigl[-1
+
\frac{1}{2N}{\rm Tr}\Bigl(\frac{D\hat{\boldsymbol{X}}_i}{Dt}\frac{D\hat{\boldsymbol{X}}_i}{Dt}
+
\frac{R_{11}^2}{2\ell_{11}^6}[\boldsymbol{X}_i,\boldsymbol{X}_j]
[\boldsymbol{X}_i,\boldsymbol{X}_j] \Bigr)+O(\frac{1}{N^2})\Bigr], 
\end{align}
as expected from the relation between the DLCQ scheme 
and the original BFSS proposal. 
Here it is assumed that both of  the kinetic term 
${\rm Tr}\Bigl(\frac{D\hat{\boldsymbol{X}}_i}{Dt}
\frac{D\hat{\boldsymbol{X}}_i}{Dt}\Bigr)$  and 
the potential term ${\rm Tr}\bigl([\boldsymbol{X}_i,\boldsymbol{X}_j]
[\boldsymbol{X}_i,\boldsymbol{X}_j]\bigr)$ are {\it at most} of order 
one. 

After these non-covariant gauge fixings, the naive Lorentz transformation 
laws expressed by (\ref{lorentzgene}) must be modified by taking into account compensating 
gauge transformations. Though we do not work out 
formalistic details along this line, it is to be noted 
that such deformed 
transformation laws are necessarily different from 
those expected from the classical theory of membranes.


\vspace{0.2cm}
\noindent
{\it Remarks}

(i) One of the novel characteristics in our model is that 
the 11 dimensional Planck length $\ell_{11}$ emerges as the 
expectation value (\ref{11dlength}) of an invariant $X_{\uhdb}^2$,  
arising
out of a completely scale-free theory. 
Together with a compactified unit $R_{11}$ (or $R$) of momentum, they provide two independent constants $g_s$ and $\ell_s$ of string theory embedded in 11 dimensions. 
This emerges once we specify a particular 
solution for $P_{\circ}^{\mu}$ and $X_{{\rm M}}^{\mu}$ as initial conditions through these conserved 
quantities. However, 
the meaning of the Lorentz invariant $X_{\uhdb}^2$ is quite different from 
 $P_{\circ}^{\mu}$. 
The former determines the coupling constant 
for the time-evolution of traceless matrix variables in a 
 Lorentz-invariant manner, 
while the latter only specifies the initial 
values of center-of-mass momentum 
which is essentially decoupled from the dynamics of 
the traceless matrix part. 
It seems natural to postulate that 
the invariant $X_{\uhdb}^2$ defines a super-selection 
rule with respect to the scale symmetry of our system. 
In other words, we demand that 
no superposition is allowed among states with 
different values of  $X_{\uhdb}^2$. 
Due to the scale symmetry, any pair of different sectors of the  
Hilbert space (after quantization) can be 
mapped into each other by an appropriate scale transformation, 
and then all the different super-selection sectors 
describe completely the same dynamics. 
In this sense, 
the scale symmetry is {\it spontaneously} broken. 
Such a fundamental nature of 11 dimensional 
gravitational length is also one of the 
expected general properties of M-theory. 

On the other hand, states with varying 
components of the vector $X_{\uhdb}^{\mu}$ connected by Lorentz transformations with a fixed $X_{{\rm M}}^2$ are not forbidden to be 
superposed, along with the center-of-mass momentum $P_{\circ}^{\mu}$. 
In fact, the $\delta_{w}$-gauge Gauss constraint (\ref{orthogonality1}) 
requires this: 
depending on the light-front foliation or time-like foliation, it leads to relations among these conserved quantities, respectively, 
\begin{align}
P_{\circ}^-
=-\frac{P_{\circ}^+}{(X_{\uhdb}^+)^2}X_{\uhdb}^2
\qquad 
\mbox{or} 
\qquad
P_{\circ}^0
=\frac{P_{\circ}^{10}X_{\uhdb}^{10}}{\sqrt{(X_{\uhdb}^{10})^2-X_{\uhdb}^2}}.
\label{lfrelation}\end{align}
Thus, given the center-of-mass ``energies",  
compactification radii and gravitational length, these relations 
determine $X_{\uhdb}^+$ or $X_{\uhdb}^{10}$. 
In particular, the light-like limit $P_{\circ}^-\rightarrow 0$ 
with finite $P_{\circ}^+$ (or $P_{\circ}^{10}$)  
corresponds to a singular limit 
$X_{\uhdb}^+\rightarrow \infty$ or equivalently to $X_{\uhdb}^{10}
\rightarrow \infty$. 

\vspace{0.2cm}
(ii)  The fact that the system is reducible 
from 11 (10 spatial and 1 time-like) matrix degrees of freedom to 
9 spatial matrix degrees of freedom is 
of course due to the presence of the higher gauge symmetries. 
From the viewpoint of ordinary 
relativistic mechanics of many particles, this feature is also quite a peculiar phenomenon: 
our higher gauge symmetries imply that two space-time 
directions corresponding to the M-plane are {\it  
locally unobservable} with respect to the dynamics of 
M-theory partons. That is the reason why we can 
eliminate both of the traceless parts, $\hat{\boldsymbol{X}}^{\pm}$ and $\hat{\boldsymbol{P}}^{\pm}$ of the matrix degrees of freedom along the 
M-plane.\footnote{In the case 
of a single string or of a single membrane, the light-front gauge $\partial_{\sigma}
X^+=0$ 
allows us to express $X^-$, as a passive variable which does not 
participate in the dynamics, in terms of 
transverse variables. In contrast, in our model, 
we can eliminate the traceless part 
$\hat{\boldsymbol{X}}^-$, and thus our 
higher gauge symmetries play a much stronger role than the 
re-parametrization invariance in string and membrane theories. 
The possibility of different formulations 
which are more analogous to strings and membrane might be 
worthwhile to pursue. However, that would require a 
framework which is different from 
the present paper.  }  
If $X_{{\rm M}}^+\hat{\boldsymbol{X}}^-$ in (\ref{xminus}) were not  eliminated in the above light-front gauge fixing, we would have $-\bigl(X_{{\rm M}}^+X^-_{ab}(x_a^i-x_b^i)\bigr)^2$ giving non-zero potential 
of {\it wrong} sign 
for purely diagonal configurations with respect to the transverse directions. 
The absence\footnote{Note also that 
the absence of this term, being of wrong sign, is required for supersymmetry. } of this term conforms to,   
at least qualitatively, one remarkable aspect of 
general-relativistic interactions 
of M-theory partons. Due to the elimination of $\hat{\boldsymbol{X}}^-$, the 
{\it static} diagonal matrices (with $\hat{\boldsymbol{P}}^i=0$ and 
$[\boldsymbol{X}_i, \boldsymbol{X}_j]=0$) for all directions 
transverse to the plane spanned by $P_{\circ}^{\mu}$ and 
$X_{\uhdb}^{\mu}$ 
provide exact classical solutions describing degenerate 
ground states with ${\cal M}_{{\rm boson}}^2=0$, corresponding to the flat directions of the potential term, whose existence is also a consequence of 
the structure of our 3-bracket. 
In classical particle pictures, this corresponds to bundles of parallel (and collinear as a special degenerate limit) trajectories of 
11 dimensional gravitons. On the other hand, 
in classical general relativity, it is well known that the parallel pencil-like trajectories 
of massless particles are non-interacting: equivalently, for the metric 
of the form
\begin{align}
ds^2=dx^{\mu}dx_{\mu}+h_{--}(dx^-)^2
\end{align}
with coordinate condition $\partial_+h_{--}=0$, 
the vacuum Einstein equations reduce to the linear Laplace 
equation $\partial_i^2 h_{--}=0$ in the transverse space around such 
trajectories \cite{aichen}. This makes possible the 
interpretation of states with higher quantized momenta $P^+_{\circ}$ 
as composite states consisting of constituent states with 
unit momentum $1/R$ along the compactified 
direction.  Note that in ordinary local theories of point-like particles, 
a state of a single particle with multiple units of momentum and 
a state of many particles of the same total momentum but with various different distributions 
of constitutent's momenta must be 
treated as different states which can be discriminated by 
relative positions in the coordinate representation. In contrast to this, our higher gauge symmetries 
render the relative positions along the $x^-$ directions unobservable as 
unphysical degrees of freedom.

\vspace{0.2cm}
(iii) As regards classical solutions with diagonal 
transverse degrees of matrices, there is another curious property 
for non-static solutions with constant non-zero velocities for finite $N$. 
The 
action (\ref{timelikeaction}) in the time-like gauge 
shows that the upper bound 
for the magnitude of transverse {\it relative} velocities is described by
\begin{align}
N\ge {\rm Tr}\Bigl(\frac{D\hat{\boldsymbol{X}}_i}{Dt}
\frac{D\hat{\boldsymbol{X}}_i}{Dt}\Bigr) 
\end{align}
For classical diagonal configurations with vanishing gauge fields, the right-hand side 
reduces to the sum of squared velocities $\sum_{a=1}^N\bigl(d\hat{X}_{aa}^i/dt)^2$, and hence for symmetric 
distributions of D-particles such that $v\equiv | d\hat{X}_{aa}^i/dt|$ is indenpendent of $a$, 
this bound corresponds to the usual relativistic bound $v\le c=1$ 
in terms of {\it absolute} (not relative) velocities. 
On the other hand, for non-symmetrical configurations, this, being a  
bound averaged over relative velocities of constituent partons 
and the off-diagonal degrees of freedom, 
does not forbid the appearance of super-luminal 
velocities for a part of constituent partons, when other partons have sub-luminal (or zero) 
velocities provided $N\ge 3$. This situation is owing to the absence of 
the mass-shell conditions set independently for each 
parton, and is actually expected in any 
covariantized extensions of the light-front 
super quantum mechanics, which itself has no 
such condition,
\footnote{
For the system of a single particle as exemplified 
in the Introduction, the relativistic upper bound 
is automatically built-in, due to the mass-shell condition. The problem only appears 
for many-body systems when the mass-shell condition 
for each particle-degree of freedom is not independently imposed. 
For comparison, if we consider a system of  $N$ free massive 
particles designated by $a=1, 2, \ldots N$ and impose mass-shell condition for each particle, the usual relativistic upper bound $|v_i^{(a)}|=
\bigl|\frac{dx^(a)_i}{dx^0}\bigr|<1$ for the transverse velocities can be expressed, 
in terms of a {\it common} light-like time $x^+=x^{10}+x^0$, 
 as 
$\Bigl|\frac{dx_i^{(a)}}{dx^+}\Bigr|^2<
\frac{1-(v_{10}^{(a)})^2}{(1+v_{10})^2}$ for each $a$ separately, 
where  $v_{10}$ in the denominator is the center-of-mass velocity 
along the 10th spatial direction whose absolute value can be fixed to be an 
arbitrary value less than 1, providing that 
the center-of-mass momentum is time-like. In terms of independent light-front times $x^{+\, 
(a)}$, the bounds are  
$\Bigl|\frac{dx_i^{(a)}}{dx^{+\, (a)}}\Bigr|^2<\frac{1-v_{10}^{(a)}}{1+v_{10}^{(a)}}$, and hence there is no 
restriction on the magnitude for transverse velocities, 
as the right-hand side can become arbitrarily large as 
$v_{10}^{(a)}\rightarrow -1$.} 
as we have already mentioned in the Introduction. 
Note, however, that the role of 
these peculiar states would be negligible in any well defined large $N$ limits of our interest. 

\section{Fermionic degrees of freedom and supersymmetry}

Our next task is to extend foregoing constructions to a supersymmetric 
theory. Since we already know a 
 supersymmetric version reduced to the light-front gauge 
with the DLCQ compactification, all we need is to find a way of 
reformulating it in terms of appropriate languages which 
fit consisitently to the structure of the previous bosonic part without 
violating covariance in the sense of 11 dimensional Minkowski space-time 
and other symmetries. 
Corresponding to the traceless part of the 
bosonic matrices, we introduce 
Majonara spinor Hermitian traceless matrices denoted by $\boldsymbol{\Theta}$. 
By this, we mean that all the {\it would-be} real components of matrix elements are 
Majonara spinors with 32 components.\footnote{
The Dirac matrices $\Gamma^{\mu}$ are in the 
Majonara real representation where all components are real numbers, and 
$\{\Gamma^{\mu},\Gamma^{\nu}\}=2\eta^{\mu\nu}, 
(\Gamma^{\mu})^{{\rm T}}=\Gamma^0\Gamma^{\mu}
\Gamma^0$, and $\Gamma^{\mu_1\mu_2\ldots \mu_n}$ is 
a totally anti-symmetrized product of $n$ matrices, 
so that $(\Gamma^0\Gamma^{\mu_1\mu_2\ldots \mu_n})^{{\rm T}}
=(-1)^{n-1}\Gamma^0\Gamma^{\mu_{n}\mu_{n-1}\ldots \mu_1}$. } The Dirac conjugate is defined by 
$\bar{\boldsymbol{\Theta}}_{ab}=\boldsymbol{\Theta}^{{\rm T}}_{ab}\Gamma^0$ where the transposition symbol T is with resect to spinor components treated as column and row vectors; but we 
mostly suppress the T-symbol on $\boldsymbol{\Theta}$ below, because
 it must be obvious by the position of Gamma 
matrices acting on them. 

To be a supersymmetric theory, we also need the fermionic partner for the 
center-of-mass degrees of bosonic variables. 
The fermionic center-of-mass degrees of freedom, being 
a single 32 component Majorana spinor, are denoted by 
$\Theta_{\circ}$ with 
the subscript $\circ$ as in the bosonic case. 
Unlike bosonic case, the relative normalization between the traceless fermion matrices and $\Theta_{\circ}$ can be chosen
arbitrarily since it is completely decoupled from the 
dynamics of the traceless matrices. We therefore treat the fermionic 
matrices  $\boldsymbol{\Theta}$ always as traceless, being completely 
separated from the center-of-mass fermionic 
variables $\Theta_{\circ}$.\footnote{For notational brevity, we drop the 
symbol `` $\, \hat{}\, $ " for fermionic matrices, as for other bosonic 
variables such as $\boldsymbol{A}, \boldsymbol{B}, \boldsymbol{Z}$ which are defined as traceless from the beginning. }  Note that 
in the bosonic case, the center-of-mass motion couples with the 
traceless part through the Hamiltonian 
constraint, although their equations of motion are decoupled. 
Under the 
$\tau$-reparametrization, both $\Theta_{\circ}$ and $\boldsymbol{\Theta}$ transform as scalar.

We aim at a minimally possible 
extension of the light-front Matrix theory. 
A fundamental premise in what follows 
is that for fermionic 
variables, there is no counterpart of the bosonic M-variables, a 
canonical (non-matrix) pair $(X_{\uhdb}^{\mu}, P_{\dhd}^{\mu})$.
This requires that the Gauss constraints (\ref{orthogonality1}) and 
(\ref{bosongauss2}) involving them 
must themselves be invariant under supersymmetry transformations. 
This will be achieved by requiring that the center-of-mass 
momentum $P_{\circ}^{\mu}$ is super invariant, 
and consequently the Gauss constraint (\ref{orthogonality2}) 
should also be super invariant. 
To be consistent with these demands, 
the fermionic variables are not subject to gauge 
transformations except for 
$\delta_{HL}\Theta=\delta_{HY}\Theta=\delta_H\Theta=(0, i\sum_r[F^r,G^r,\Theta])$, 
which is 
reduced simply {\it only} to the usual SU($N$) gauge transformation 
corresponding to the gauge field $\boldsymbol{A}$, 
\begin{align}
\delta_{H}\boldsymbol{\Theta}=i[\boldsymbol{H}, \boldsymbol{\Theta}].
\end{align}
Consequently the usual traces of the products of fermion matrices give 
gauge invariants, provided they do not involve bosonic 
matrix variables, while the products involving both fermionic and 
bosonic matrices can be made invariant by combining them into 
3-brackets, just as in the case of purely bosonic 
cases. 
Since the fermionic variables  intrinsically obey the first-order formalism 
in which the generalized coordinates and momenta are 
mixed inextricably among spinor components and hence 
the fermionic generalized coordinates and 
momenta should have the same transformation laws, 
it would be very difficult to extend the structure of higher-gauge transformations 
for the bosonic variables to fermionic variables covariantly 
if we assumed non-zero fermionic M-variables. But that is not 
necessary as we shall argue below. 

\subsection{Center-of-mass part: 11 dimensional rigid supersymmetry}

Let us now start from the center-of-mass degrees of freedom. 
Since we require that the theory has at least 11 dimensional 
rigid supersymmetry, it is natural to set the 
center-of-mass part in a standard fashion as for 
the case of a single point particle. 
Thus the fermionic action is chosen to be 
\begin{align}
\int d\tau P_{\circ\, \mu}\bar{\Theta}_{\circ}\Gamma^{\mu}
\frac{d\Theta_{\circ}}{d\tau}, 
\end{align}
which is obtained by making a replacement $\frac{dX^{\mu}_{\circ}}{d\tau}
\rightarrow \frac{dX^{\mu}_{\circ}}{d\tau}+
\bar{\Theta}_{\circ}\Gamma^{\mu}
\frac{d\Theta_{\circ}}{d\tau}$  from the 
center-of-mass part of the bosonic Poincar\'{e} integral.
 Under the usual rigid super translation 
\begin{align}
\delta_{\varepsilon}\Theta_{\circ}=-\varepsilon, 
\end{align}
together with the 
requirement
\begin{align}
\delta_{\varepsilon}P_{\circ}^{\mu}=0, 
\end{align}
the action is invariant by assuming the transformation law 
for the bosonic center-of-mass coordinates as 
\begin{align}
\delta_{\varepsilon}X_{\circ}^{\mu}=\bar{\varepsilon}
\Gamma^{\mu}\Theta_{\circ},
\label{rigidsuper}
\end{align}
since 
\begin{align}
\delta_{\varepsilon}\Bigl(  \frac{dX^{\mu}_{\circ}}{d\tau}+
\bar{\Theta}_{\circ}\Gamma^{\mu}
\frac{d\Theta_{\circ}}{d\tau}\Bigr)=0, 
\label{xthetainv}
\end{align}
which is consistent with the first order equations of motion.

Under the assumption that all the other variables not exhibited above are 
inert with respect to the rigid super transformation, it is clear that  the existence of  these 
fermionic center-of-mass degrees of 
freedom does not spoil any of symmetry properties introduced 
in previous sections, 
{\it provided} that the remaining matrix part of the action 
decouples from $X_{\circ}^{\mu}, \Theta_{\circ}$ and $P_{{\rm M}}^{\mu}$. 
This ensures 
 that the first-order equations of motion for the canonical 
pairs $(X_{\circ}^{\mu}, P_{\circ}^{\mu})$ and $(X_{{\rm M}}^{\mu}, 
P_{{\rm M}}^{\mu})$  
are of the following form, reflecting conservation laws and the 
passive nature of the associated cyclic variables, 
\begin{align}
&\frac{dP_{\circ}^{\mu}}{d\tau}=0, \\
&\frac{1}{e}\Bigl(
\frac{DX_{\circ}^{\mu}}{D\tau}+\bar{\Theta}_{\circ}\Gamma^{\mu}
\frac{d\Theta_{\circ}}{d\tau}\Bigr)=P_{\circ}^{\mu}-f^{\mu}, 
\label{passivex}\\
&\frac{dX_{{\rm M}}^{\mu}}{d\tau}=0, \\
&\frac{1}{e}\frac{DP_{{\rm M}}^{\mu}}{D\tau}=g^{\mu}, 
\label{passivepm}
\end{align} 
where the unspecified functions $f^{\mu}$ and $g^{\mu}$ are  contributions from the remaining 
part of action 
and do not depend on these passive variables themselves. 
It should also be mentioned that the scale dimensions of the 
fermion center-of-mass variables are 
\begin{align}
\Theta_{\circ}\rightarrow \lambda^{1/2}\Theta_{\circ}, 
\quad 
\varepsilon\rightarrow \lambda^{1/2}\varepsilon.  
\end{align}

The 
equation of motion for the fermionic center-of-mass spinor is then
\begin{align}
P_{\circ}\cdot \Gamma\frac{d\Theta_{\circ}}{d\tau}=0.
\end{align}
For generic case with non-vanishing effective mass square 
$-P_{\circ}^2>0$, 
this leads to a 
conservation law
\begin{align}
\frac{d\Theta_{\circ}}{d\tau}=0. 
\end{align}
In general, the quantum states consist of fundamental massive 
super-multiplets of dimension $2^{16}$. 

We here briefly touch the canonical structure of the fermionic 
center-of-mass variables. 
From the above action, there is a primary second-class constraint, 
\begin{align}
\Pi_{\circ}+\bar{\Theta}P_{\circ}\cdot\Gamma=0, 
\label{c-o-m-fermiprimary}
\end{align}
satisfying a Poisson bracket relation
\begin{align}
\{\Pi_{\circ\, \alpha}+(\bar{\Theta}P_{\circ}\cdot\Gamma)_{\alpha}, 
\Pi_{\circ\, \beta}+(\bar{\Theta}P_{\circ}\cdot\Gamma)_{\beta}\}_{{
\rm P}}
=2(\Gamma^0P_{\circ}\cdot \Gamma)_{\alpha\beta}, 
\end{align}
where $\Pi_{\circ}$ is canonically conjugate to 
$\Theta_{\circ}$ and $\alpha, \beta, \ldots$ are spinor indices. 
Correspondingly, the Poisson bracket must be replaced by Dirac bracket, which is also required to render the canonical 
structure supersymmetric. 
We give a brief account of this topic in appendix A. 

In the limit of light-like center-of-mass momentum $P_{\circ}^2=0$, a one-half of the primary constraints (\ref{c-o-m-fermiprimary}) becomes first class because of the 
existence of zero eigenvalues for the Dirac operator 
$P_{\circ}\cdot \Gamma$, and the fermionic 
 equations of motion have a redundancy. 
In the present work, 
we will not elaborate on remedying this complication, 
by assuming generic massive case. 
Physically, this is allowed since the system, describing a general 
many-body 
system with massless gravitons, has 
continuous mass spectrum {\it without} mass gap. 
When we have to deal with 
the light-like case, we can always consider a slightly 
different state with a small but non-zero center-of-mass 
by 
adding soft gravitons propagating with a non-zero small 
momentum along directions transverse to the 
original states. 

As is well known, the singularity at $P_{\circ}^2=0$ is 
associated with the emergence of a
 local symmetry, called Siegel (or ``$\kappa$"-) symmetry \cite{siegel}, 
\begin{align}
\delta_{\kappa}\Theta_{\circ}=P_{\circ}\cdot \Gamma \kappa, \quad 
\delta_{\kappa}X_{\circ}^{\mu}=-\bar{\Theta}_{\circ}\Gamma^{\mu}
\delta_{\kappa}\Theta_{\circ}, 
\end{align}
with arbitrary spinor function $\kappa(\tau)$.\footnote{
The action is invariant, under the condition 
${\cal M}^2=0$ (which holds identically in the trivial case $N=1$), by adjoining the transformation of 
ein-bain 
$\delta_{\kappa}e=-4\frac{d\bar{\Theta}_{\circ}}{d\tau}\kappa$. 
Of course, the expression of the 
effective mass square is to be extended by including 
the contribution of traceless fermionic matrices, as discussed below. 
}
This allows us to eliminate a half of components of $\Theta_{\circ}$ 
by a suitable redefinition of $X_{\circ}^{\mu}$, and hence 
the super-multiplets are shorten to $2^{16/2}=2^8=256$ dimensions 
(or to half-BPS states). This coincides with the 
dimension of graviton super-multiplet in 11 dimensions which 
constitutes the basic physical field-degrees of freedom of 11 dimensional 
supergravity. It should be noted, however, that 
generic many-body states with time-like 
center-of-mass momenta composed of 
massless short multiplets obey ``longer" massive representations. 
For instance, a generic two-body {\it scattering} state of gravitons with 
$-P_{\circ}^2>0$ would constitute a massive 
multiplet of $2^8\times 2^8=2^{16}$ dimensions. 
Therefore, 
it does not seem
 reasonable to demand a $\kappa$-symmetry as a general condition 
in our case of the center-of-mass supersymmetry, since we are dealing with 
$N=1$ supersymmetry in the highest 11 dimensions.\footnote{Note that the situation is different for a single supermembrane in 11 dimensions, where the ground state is required to be a massless graviton supermultiplet. 
 It is also to be mentioned that in lower space-time dimensions the 
$\kappa$-symmetry can be generalized to massive case when we have 
an extended supersymmetry with non-vanishing central charges. 
See {\it e.g.} \cite{evans}. 
This is consistent 
with the fact that such systems can be obtained by 
dimensional reduction from massless theories of higher dimensions, 
by which massive states can constitute a short multiplet with respect to 
extended supersymmetries. 

}

\subsection{Traceless matrix part: dynamical supersymmetry}

Next, we proceed to the traceless matrix part. A natural candidate for 
the transformation law of the bosonic matrices is 
\begin{align}
\delta_{\epsilon}\hat{\boldsymbol{X}}^{\mu}=
\bar{\epsilon}\Gamma^{\mu}\boldsymbol{\Theta}.
\label{tracesusy}
\end{align}
Superficially the previous 
transformation (\ref{rigidsuper}) may be regarded as the 
trace part of this form, but we will shortly see critical differences. 
To keep the difference in mind, the spinor parameter is now denoted 
by a symbol $\epsilon$ which is distinct from that ($\varepsilon$) for the 
center-of-mass degrees of freedom, since they are in principle 
independent of each other and can be treated separately. 
This is natural, since the traceless matrices describe the 
internal dynamics of relative degrees of freedom. 
Following common usage, we call the rigid supersymmetry of the center-of-mass part 
``kinematical" which is essentially a superspace translation as 
a partner of rigid space-time translation, and that of the traceless part ``dynamical",   
mixing between the bosonic and 
fermionic traceless matrices 
without any inhomogeneous 
shift-type contributions. The dynamical supersymmetry 
of our system will be related to rigid translations with respect to 
the invariant time parameter $s$ ($ds=ed\tau$). 
Once these two independent 
supersymmetries are established, however, we can 
combine them depending on different situations. 
For instance, we can partially identify $\epsilon$ and 
$\varepsilon$ up to some proportional factor and projection 
(or twisting) 
conditions with respect to spinor indices. That would occur through an identification of the invariant 
proper-time parameter with an external time coordinate as 
a gauge choice for re-parametrization invariance, as 
in the case of the usual formulation of 
the light-front Matrix theory.

\vspace{0.2cm}
\noindent
(1) {\it Projection conditions}

In discussing the transformation law for $\boldsymbol{\Theta}$, 
we have to take into account the existence of the Gauss 
constraint (\ref{orthogonality1}) which characterizes the M-plane. 
We treat this constraint as a strong constraint in studying dynamical supersymmetry.
This is allowed, as long as Lorentz covariance 
is not lost.  
We then have to assume the 
equations of motion for the center-of-mass part and for 
the M-variables strongly, so that 
we can use the conservation laws of $P_{\circ}^{\mu}$ and 
$X_{\uhdb}^{\mu}$, both of which are assumed to be 
inert $\delta_{\epsilon}P^{\mu}_{\circ}=0=
\delta_{\epsilon}X^{\mu}_{{\rm M}}$ against dynamical as 
well as kinematical super transformations. 
We do not expect any difficulty with 
this restriction at least {\it practically}: for example, 
we can use the representation 
where both of these vectors are diagonalized for 
quantization. 
Thus it should be kept in mind that 
the supersymmetry transformation 
laws derived below have validity only ``on shell" with respect to these variables.  With respect to
 the traceless matrix part, on the other hand, they will be valid without 
 using the equations of motion.

Now we have to examine the compatibility of the other Gauss constraints 
(\ref{orthogonality2}) and (\ref{bosongauss2}) with dynamical supersymmetry. 
Our assumptions, with the dynamical 
super transformation 
(\ref{tracesusy}), requires that $\delta_{\epsilon}(P_{\circ}\cdot
\hat{\boldsymbol{X}})=0$, namely, 
\begin{align}
\bar{\epsilon}P_{\circ}\cdot \Gamma\boldsymbol{\Theta}=0. 
\label{psupercond}
\end{align}
It is also necessary to demand $\delta_{\epsilon}(X_{{\rm M}}
\cdot \hat{\boldsymbol{P}})=0$ for the 
momentum as 
\begin{align}
X_{{\rm M}}\cdot \delta_{\epsilon}\hat{\boldsymbol{P}}=0. 
\label{xmsupercond}
\end{align}
We first concentrate on the former. In any 
natural decomposition between generalized 
coordinates and momenta for the spinor 
components of $\boldsymbol{\Theta}$, this is a second-class constraint. 
This suggests that the traceless spinor matrix and parameter 
$\epsilon$ should obey certain projection condition strongly, rather than as a Gauss constraint 
associated with gauge symmetry, such that 
(\ref{psupercond}) is obeyed. 
By the existence of two conserved vectors $P_{\circ}^{\mu}$ and 
$X_{\uhdb}^{\mu}$ which are orthogonal to 
each other due to the strong constraint (\ref{orthogonality1}), we have a candidate for 
Lorentz-invariant (real) projector:
\begin{align}
P_{\pm}\equiv \frac{1}{2}(1\pm \Gamma_{\circ}\Gamma_{\uhdb}).
\end{align}
Here we have introduced 
\begin{align}
\Gamma_{\uhdb}\equiv \frac{X_{\uhdb}\cdot \Gamma}{\sqrt{X_{\uhdb}^2}}, 
\quad 
\Gamma_{\circ}\equiv \frac{P_{\circ}\cdot \Gamma}{
\sqrt{-P_{\circ}^2}},
\end{align}
by assuming generic cases with time-like center-of-mass momentum
 $-P_{\circ}^2>0$ as before. 
Due to the orthogonality constraint (\ref{orthogonality1}), 
these Lorentz-invariant Dirac matrices satisfy
\begin{align}
\Gamma_{\uhdb}\Gamma_{\circ}+\Gamma_{\circ}\Gamma_{\uhdb}=0, 
\quad 
\Gamma_{\uhdb}^2=1, \quad \Gamma_{\circ}^2=-1, 
\quad 
(\Gamma_{\circ}\Gamma_{\uhdb})^2=1, 
\end{align}
and consequently
\begin{align}
\Gamma_{\uhdb}(\Gamma_{\circ}\Gamma_{\uhdb})&=
-(\Gamma_{\circ}\Gamma_{\uhdb})\Gamma_{\uhdb}, 
\quad 
\Gamma_{\circ}(\Gamma_{\circ}\Gamma_{\uhdb})=
-(\Gamma_{\circ}\Gamma_{\uhdb})\Gamma_{\circ}, \\
& P_+\Gamma_{\uhdb}=\Gamma_{\uhdb}P_-, \,\, 
P_+\Gamma_{\circ}=\Gamma_{\circ}P_-, \\
&P_{\pm}^2=P_{\pm}, \, P_{\pm}P_{\mp}=0.
\end{align}
Note that 
\begin{align}
P_{\pm}\Gamma_i=
\Gamma_iP_{\pm}
\end{align}
for the SO(9) directions $i$, transverse to the M-plane.\footnote{There is another possible projector $\tilde{P}_{\pm}\equiv \frac{1}{2}(1\pm  \Gamma_{\uhdb})$. However this does not discriminate the directions of $P_{\circ}^{\mu}$ from the other SO(9) space-like directions, and 
is not suitable for our purpose here.}
 
 We then 
introduce the projection condition by $\Gamma_{\circ}\Gamma_{\uhdb}
\boldsymbol{\Theta}=-\boldsymbol{\Theta}$, namely, 
\begin{align}
P_-\boldsymbol{\Theta}=\boldsymbol{\Theta}, \, \, P_+\boldsymbol{\Theta}=0, \quad 
\mbox{(or \,\, equivalently \,\, $\bar{\boldsymbol{\Theta}}P_+=
\bar{\Theta},\, \, \bar{\boldsymbol{\Theta}}P_-=0$)} 
\label{projection}
\end{align}
together with the opposite projection on $\epsilon$, 
\begin{align}
P_+\epsilon=\epsilon, \, \, P_-\epsilon=0, 
\quad 
\mbox{(or \,\, equivalently \,\, $\bar{\epsilon}P_-=
\bar{\epsilon},\, \, \bar{\epsilon}P_+=0$)} .
\label{projection2}
\end{align}
Then as desired 
\begin{align}
\bar{\epsilon}P_{\circ}\cdot \Gamma\boldsymbol{\Theta}
=\bar{\epsilon}P_-(P_{\circ}\cdot \Gamma) P_-
\boldsymbol{\Theta}
=\bar{\epsilon}(P_{\circ}\cdot \Gamma) P_+P_-\boldsymbol{\Theta}=0,
\end{align}
and simultaneously we also have, 
\begin{align}
\bar{\epsilon}X_{\uhdb}\cdot \Gamma\boldsymbol{\Theta}
=\bar{\epsilon}P_-(X_{\uhdb}\cdot \Gamma)P_-\boldsymbol{\Theta}=\bar{\epsilon}X_{\uhdb}\cdot \Gamma P_+P_-\boldsymbol{\Theta}=0,
\label{Mspinorcond}
\end{align}
while 
\begin{align}
\bar{\epsilon}\Gamma_i\boldsymbol{\Theta}=
\bar{\epsilon}P_-\Gamma_iP_-\boldsymbol{\Theta}
=\bar{\epsilon}\Gamma_iP_-\boldsymbol{\Theta}
=\bar{\epsilon}P_-\Gamma_i\boldsymbol{\Theta}
\end{align}
can be non-vanishing 
for all $i$'s, transverse to both $P_{\circ}$ and $X_{\uhdb}$. 
The dynamical supersymmetry is thus effective essentially in the 
directions which are transverse to the M-plane, 
in conformity with our requirement. 
This automatically ensures the remaining requirement (\ref{xmsupercond}), as 
we will confirm later. 

It is to be noted that the condition (\ref{projection}) is 
equivalent to 
\begin{align}
(\Gamma_{\circ}-\Gamma_{\uhdb})\boldsymbol{\Theta}=0,
\label{covlightcone}
\end{align}
which can be regarded as a Lorentz-covariant 
version of a familiar light-front gauge condition
$\Gamma^+\boldsymbol{\Theta}=0$. 
In fact, using the light-front frame defined in the previous section, 
we can rewrite (\ref{covlightcone}) using (\ref{lfrelation}) as 
($\Gamma^{\pm}=\Gamma^{10}\pm \Gamma^0$)
\begin{align}
0
&=\frac{1}{2\sqrt{-P_{\circ}^2}}\Bigl(
P_{\circ}^+\Gamma^-+
P_{\circ}^-\Gamma^+
-
\frac{\sqrt{-P_{\circ}^+P_{\circ}^-}}{\sqrt{X_{\uhdb}^2}}
(X_{\uhdb}^+\Gamma^-+X_{\uhdb}^-\Gamma^+)
\Bigr)\boldsymbol{\Theta}\nonumber \\
&=\frac{1}{2\sqrt{-P_{\circ}^2}}\Bigl(
P_{\circ}^+\Gamma^-+
P_{\circ}^-\Gamma^+
-
\frac{P_{\circ}^+}{X_{\uhdb}^+}
(X_{\uhdb}^+\Gamma^-+\frac{X_{\uhdb}^2}{X_{\uhdb}^+}\Gamma^+)
\Bigr)\boldsymbol{\Theta}=-\sqrt{-
\frac{P_{\circ}^-}{P_{\circ}^+}}\Gamma^+\boldsymbol{\Theta}. 
\end{align}
In the classical theory of a single supermembrane, 
the possibility of 
a similar projection owes to the existence of 
the $\kappa$-symmetry. 
In our system, by contrast,  the existence of the gauge-invariant Gauss constraints in the bosonic sector, involving 
dynamical variables without fermionic partners, requires 
us, on our premise of a minimal extension,  
necessarily to introduce 
projection condition for fermionic variables in a Lorentz-covariant and gauge-invariant manner. Thus our strategy can be different\footnote{
This does not exclude the possibility of 
introducing the fermionic partner even for 
the M-variables in conjunction with some 
higher {\it fermionic} gauge symmetries. 
It does not seem however that elaboration toward such a {\it non}-minimal extension is practically useful. 
}: 
we need not bother about possible 
imposition of a generalized $\kappa$-like symmetry 
for traceless matrix variables. 
   The dynamical supersymmetry 
requires that the physical degrees of freedom of traceless matrices match between bosonic and fermionic variables. 
On the bosonic side, the number of physical degrees of freedom after imposing all 
contraints is 8, counting the pairs of canonical 
variables, if we take into account all of the 
Gauss constraints including the $\boldsymbol{A}$-gauge 
symmetry. 
The number of physical degrees of freedom for the fermionic traceless 
matrices must therefore be 16, and this was made possible by our {\it covariant}  projection condition (\ref{projection}) as 
a partner of the bosonic 
constraints represented by the set of Gauss constraints, 
thanks to the existence of the M-variables.

\vspace{0.2cm}
\noindent
(2) {\it Fermion action and dynamical supersymmetry transformations}

We are now ready to present 
 the fermionic part of the action and supersymmetry 
transformations. 
The total fermionic contribution to be added to the bosonic action (\ref{bosonaction}) is 
\begin{align}
A_{{\rm fermion}}=
\int d\tau \, \Bigl[\bar{\Theta}_{\circ}P_{\circ}\cdot \Gamma \frac{d\Theta_{\circ}}{d\tau}
+\frac{1}{2}{\rm Tr}\Bigl(
\bar{\boldsymbol{\Theta}}\Gamma_{\circ}
\frac{D\boldsymbol{\Theta}}{D\tau}\Bigr)
-e\frac{i}{4}\bigl<
\bar{\Theta}, \Gamma_{\mu\nu}[X^{\mu}, X^{\nu}, \Theta]\bigr>\Bigr].
\label{fermiaction}
\end{align}
In the matrix form, the 3-bracket in the fermionic potential 
term is equal to 
\begin{align}
\bigl<
\bar{\Theta}, \Gamma_{\mu\nu}[X^{\mu}, X^{\nu}, \Theta]\bigr>
&=2\sqrt{X_{{\rm M}}^2}{\rm Tr}\bigl(
\bar{\boldsymbol{\Theta}}\Gamma_{\circ}\Gamma_i
[\boldsymbol{X}_i,\boldsymbol{\Theta}]
\bigr)=2\sqrt{X_{{\rm M}}^2}{\rm Tr}\bigl(
\bar{\boldsymbol{\Theta}}\Gamma_{\circ}\Gamma_{\mu}
[\boldsymbol{X}^{\mu},\boldsymbol{\Theta}]
\bigr), \label{fermipotential}
\end{align}
due to the projection condition\footnote{
Note that $\bigl<
\bar{\Theta}, \Gamma_{\mu\nu}[X^{\mu}, X^{\nu}, \Theta]\bigr>=2{\rm Tr}\bigl(\bar{\boldsymbol{\Theta}}\Gamma_{\mu\nu}X_{\uhdb}^{\mu}[\boldsymbol{X}^{\nu},\boldsymbol{\Theta}]\bigr)
$, which is rewritten as (\ref{fermipotential}) using $\Gamma_{{\rm M}}\boldsymbol{\Theta}=
\Gamma_{\circ}\boldsymbol{\Theta}$. } and 
the fact that no M-variables are associated with 
fermionic matrices. A consequence of this 
is that, due to the fermion projection condition,  
(\ref{fermipotential}) depends on the 
coordinate matrices only of directions transverse to 
the M-plane.

It is to be noted here that the traceless fermion matrices have zero scaling dimensions, 
with the dimension of 
$\epsilon$ being 1 correspondingly, in contrast to the 
case of center-of-mass fermion 
variables $\Theta_{\circ}$ and $\varepsilon$ whose scale dimensions are 
both 1/2.
This convention is convenient here 
to simplify some of the expressions,\footnote{If we like, 
we can recover the same scaling dimension for the traceless part as 
the center-of-mass side, by redefining 
$\boldsymbol{\Theta}\rightarrow (-2P_{\circ}^2)^{1/4}\boldsymbol{\Theta}, 
\epsilon\rightarrow (-2P_{\circ}^2)^{-1/4}\epsilon$.  } and no inconsistency arises as noticed before, since there is 
no coupling between $\Theta_{\circ}$ and $\boldsymbol{\Theta}$, 
and the kinematical supersymmetry transformation of the latter can be 
discussed independently of the former dynamical supersymmetry.

The dynamical supersymmetry transformations for 
matrix variables are, with the projection conditions 
(\ref{projection}), (\ref{projection2}) and the Gauss constraint (\ref{orthogonality1}) for the $\delta_{w}$-gauge symmetry, 
\begin{align}
&\delta_{\epsilon}\hat{\boldsymbol{X}}^{\mu}=\bar\epsilon\Gamma^{\mu}
\boldsymbol{\Theta},\\
&\delta_{\epsilon}\hat{\boldsymbol{P}}_{\mu}=
i\sqrt{X_{\uhdb}^2}\, \bigl[\bar{\boldsymbol{\Theta}}\Gamma_{\mu\nu}\epsilon, \tilde{\boldsymbol{X}}^{\nu}], \quad 
\delta_{\epsilon}\boldsymbol{K}=0, 
\\
&\delta_{\epsilon}\boldsymbol{\Theta}=P_-\bigl(\Gamma_{\circ}
\Gamma_{\mu}\hat{\boldsymbol{P}}^{\mu}\epsilon
-\frac{i}{2}\sqrt{X_{\uhdb}^2}\, \Gamma_{\circ}\Gamma_{\mu\nu}
\epsilon[\tilde{\boldsymbol{X}}^{\mu},\tilde{\boldsymbol{X}}^{\nu}]\bigr), \\ 
&\delta_{\epsilon}\boldsymbol{A}=\sqrt{X_{\uhdb}^2}\, \bar{\boldsymbol{\Theta}}\epsilon, \\
&
\delta_{\epsilon}\boldsymbol{B}=i\bigl(X_{\uhdb}^2\bigr)^{-1}[\delta_{\epsilon}\boldsymbol{A}, 
X_{\uhdb}\cdot \boldsymbol{X}], 
\\
&\delta_{\epsilon}\boldsymbol{Z}=i(P_{\circ}^2)^{-1}[\delta_{\epsilon}\boldsymbol{A}, P_{\circ}\cdot \boldsymbol{P}]
+\frac{X_{{\rm M}}^2}{2P_{\circ}^2}
([\delta_{\epsilon}\boldsymbol{X}^{\mu}, [P_{\circ}\cdot \boldsymbol{X}, 
\boldsymbol{X}_{\mu}]]+
[\boldsymbol{X}^{\mu}, [P_{\circ}\cdot \boldsymbol{X}, 
\delta_{\epsilon}\boldsymbol{X}_{\mu}]])
\end{align}
with
\begin{align}
&\tilde{\boldsymbol{X}}^{\mu}=
\boldsymbol{X}^{\mu}-
\frac{1}{X_{\uhdb}^2}X_{\uhdb}^{\mu}
(\boldsymbol{X}\cdot X_{\uhdb} )
-\frac{1}{P_{\circ}^2}P_{\circ}^{\mu}(\boldsymbol{X}\cdot 
P_{\circ}).
\end{align}
It is easy to check that due to our projection condition, 
(\ref{xmsupercond}) is satisfied as promised before. 
Remember again that, as we have emphasized, the equations of motion 
for the center-of-mass variables and the M-variables, especially 
conservation laws of $P_{\circ}^{\mu}$ and $X_{{\rm M}}^{\mu}$ which are 
completely inert against supersymmetry 
transformations as well as gauge transformaionts, are 
assumed here. On the other hand, the behavior of their conjugates, namely the passive 
variables, are fixed by the first order equations of motion. 
It is also to be noted that these transformation laws are 
independent of the ein-bein $e$. This implies 
that the part of the action involving $\tau$-derivatives and 
the remaining part (essentially Hamiltonian ${\cal H}$) including contributions 
with gauge fields, which does not 
involve the $\tau$-derivatives being proportional to the ein-bein $e$ 
are separately 
invariant under the supersymmetry transformations. 
This is one of the merits of the 
first-order formalism. 
A derivation of these results will be found in 
appendix B. 

In order to express the properties of these transformation laws 
from the viewpoint of canonical formalism, we need Dirac bracket. 
Here for simplicity, we take account only the 
fermionic second-class constraint for traceless fermionic variables. 
 With $\boldsymbol{\Pi}$ being the 
canonical conjugate to $\boldsymbol{\Theta}$, the primary second-class constraint for the traceless fermion matrices is
\begin{align}
\boldsymbol{\Pi}+\frac{1}{2}\bar{\boldsymbol{\Theta}}\Gamma_{\circ}=0, 
\quad (\boldsymbol{\Pi}P_-=\boldsymbol{\Pi})
\end{align}
satisfying the Poisson bracket algebra expressed in a component form\footnote{Note that 
$\{\Pi_{\alpha}^A, \Theta_{\beta}^B\}_{{\rm P}}=(P_-)_{\beta\alpha}\delta^{AB}$. 
Then, $\{\Pi_{\alpha}^A, (\bar{\Theta}^B\Gamma_{\circ})_{\beta}\}_{{\rm P}}=\delta^{AB}(P_-)_{\gamma\alpha}(\Gamma^0\Gamma_{\circ})_{\gamma\beta}=\delta^{AB}
(\Gamma^0\Gamma_{\circ}P_-)_{\beta\alpha}=\delta^{AB}(P_-^{{\rm T}}\Gamma^0\Gamma_{\circ})_{\beta\alpha}
$, due to $(\Gamma^0\Gamma_{\circ})_{\beta\alpha}=(\Gamma^0\Gamma_{\circ})_{\alpha\beta}$. }
\begin{align}
\{\Pi_{\alpha}^A+\frac{1}{2}
(\bar{\Theta}^A\Gamma_{\circ})_{\alpha}, 
\Pi_{\beta}^B+\frac{1}{2}
(\bar{\Theta}^A\Gamma_{\circ})_{\beta}\}_{{\rm P}}
=(\Gamma^0\Gamma_{\circ}P_-)_{\alpha\beta}\delta^{AB}, 
\end{align}
where we have denoted the spinor indices by $\alpha, \beta, \ldots, $.  
The indices $A, B, \ldots$ refer to the components 
with respect to the
 traceless spinor matrices using an hermitian orthogonal basis 
$\boldsymbol{\Theta}=\sum_A\Theta^A\boldsymbol{T}^A$ 
satisfying ${\rm Tr}(\boldsymbol{T}^B\boldsymbol{T}^B)=\delta^{AB}
$ of SU($N$) algebra. 
The non-trivial Dirac brackets for traceless 
matrices are then 
\begin{align}
&
\{\Theta^A_{\alpha}, 
\bar{\Theta}^B_{\beta}\}_{{\rm D}}=-(P_-\Gamma_{\circ})_{\alpha\beta}
\delta^{AB}, 
\quad 
\\
&\{\hat{X}^A_{\mu}, \hat{P}^B_{\nu}\}_{{\rm D}}=\eta_{\mu\nu}\delta^{AB}.
\end{align}
The imposition of our projection condition with respect to 
spinor indices does not cause difficulty here, since 
the symplectic structure can be consistently 
preserved within the projected space of spinors as
\begin{align}
\Gamma_{\circ}\Gamma^0P_-^{{\rm T}}=P_-\Gamma_{\circ}\Gamma^0.
\end{align}

Then we can derive 
\begin{align}
\{\bar{\epsilon}{\cal Q}, \hat{\boldsymbol{X}}^{\mu}\}_{{\rm D}}&=
-\bar{\epsilon}\Gamma^{\mu}\boldsymbol{\Theta}, \\
\{\bar{\epsilon}{\cal Q}, \hat{\boldsymbol{P}}_{\mu}\}_{{\rm D}}
&=-i\sqrt{X_{{\rm M}}^2}[
\bar{\epsilon}
\Gamma_{\mu\nu}\boldsymbol{\Theta}, \tilde{\boldsymbol{X}}^{\nu}]=i
\sqrt{X_{{\rm M}}^2}[\bar{\boldsymbol{\Theta}}\Gamma_{\mu\nu}\epsilon, 
\tilde{\boldsymbol{X}}^{\nu}],\\
\{\bar{\epsilon}{\cal Q}, \boldsymbol{\Theta}\}_{{\rm D}}&=
-P_-\bigl(\Gamma_{\circ}
\Gamma_{\mu}\hat{\boldsymbol{P}}^{\mu}\epsilon
-\frac{i}{2}\sqrt{X_{\uhdb}^2}\, \Gamma_{\circ}\Gamma_{\mu\nu}
\epsilon[\tilde{\boldsymbol{X}}^{\mu},\tilde{\boldsymbol{X}}^{\nu}]\bigr), 
\end{align}
where the supercharge is 
\begin{align}
{\cal Q}=P_-{\rm Tr}(\tilde{\hat{\boldsymbol{P}}}_{\mu}\Gamma^{\mu}\boldsymbol{\Theta}
-\frac{i}{2}\sqrt{X_{{\rm M}}^2}[\tilde{\boldsymbol{X}}^{\mu}, \tilde{\boldsymbol{X}}^{\nu}]\Gamma_{\mu\nu}\boldsymbol{\Theta})
\end{align}
with 
\begin{align}
\tilde{\hat{\boldsymbol{P}}}^{\mu}=
\hat{\boldsymbol{P}}^{\mu}-
\frac{1}{X_{\uhdb}^2}X_{\uhdb}^{\mu}
(\hat{\boldsymbol{P}}\cdot X_{\uhdb} )
-\frac{1}{P_{\circ}^2}P_{\circ}^{\mu}(\hat{\boldsymbol{P}}\cdot 
P_{\circ}).\end{align}
The supercharge satisfies\footnote{Here $[\quad, \quad ]_+$ is the 
matrix anti-commutator. The simplest way of 
checking this algebra is to go to the special frame 
introduced in the Appendix B and use the following identy \cite{banksetal} for 
$\Gamma^0\Gamma_{\circ}\Gamma^iP_-\Rightarrow \gamma^i, \quad 
\Gamma^0\Gamma_{\circ}\Gamma^{ij}P_-\Rightarrow \gamma^{ij}$, 
\[
\gamma^i_{\beta\beta'}\gamma^{ij}_{\alpha\alpha'}
+\gamma^i_{\alpha\beta'}\gamma^{ij}_{\beta\alpha'}
+\gamma^i_{\alpha\alpha'}\gamma^{ij}_{\beta\beta'}
+\gamma^i_{\beta\alpha'}\gamma^{ij}_{\alpha\beta'}=2(\gamma_{\alpha'\beta'}^j\delta_{\alpha\beta}-
\gamma^j_{\alpha\beta}\delta_{\alpha'\beta'}).
\]
Note that $\gamma_i^T=\gamma_i, \gamma_{ij}^T=-\gamma_{ij}$ 
in the projected space of spinors. 
}
\begin{align}
\{\bar{\epsilon}_1{\cal Q}, \bar{\epsilon}_2{\cal Q}\}_{{\rm D}}
=&-2(\bar{\epsilon_1}\Gamma_{\circ}\epsilon_2)
{\rm Tr}\Bigl(
\frac{1}{2}\tilde{\hat{\boldsymbol{P}}}^2-
\frac{1}{4}X_{{\rm M}}^2[\tilde{\boldsymbol{X}}^{\mu}, \tilde{\boldsymbol{X}}^{\nu}][\tilde{\boldsymbol{X}}_{\mu}, \tilde{\boldsymbol{X}}_{\nu}]
+\frac{i}{2}\sqrt{X_{{\rm M}}^2}(\bar{\boldsymbol{\Theta}}\Gamma_{\circ}\Gamma_{\mu}[\boldsymbol{X}^{\mu}, \boldsymbol{\Theta}])\Bigr)\nonumber \\
&+2(\bar{\epsilon}_1\Gamma_{\circ}\Gamma_{\mu}\epsilon_2)\sqrt{X_{{\rm M}}^2}{\rm Tr}
\bigl(
i\tilde{\boldsymbol{X}}^{\mu}[ \tilde{\boldsymbol{X}}^{\nu},\tilde{\boldsymbol{P}}_{\nu}]-\frac{1}{2}i\tilde{\boldsymbol{X}}^{\mu}[\boldsymbol{\Theta}, \Gamma^0\Gamma_{\circ}\boldsymbol{\Theta}]_+\bigr), 
\label{susyalgebra}
\end{align}
which is the covariantized version of the supersymmetry algebra (with 
{\it finite} $N$) in the usual light-front formulation. 
Note that the second line of (\ref{susyalgebra}) represents a field-dependent $\boldsymbol{A}$-gauge transformation, reflecting the 
fact that the dynamical supersymmetry transformation 
intrinsically involves an $\boldsymbol{A}$-gauge transformation. 
Thus,  up to a field-dependent 
gauge transformation, 
the commutator $[\delta_{\epsilon_1},\delta_{\epsilon_2}]$ 
induces an infinitesimal translation with respect to the invariant time parameter $s$, 
\begin{align}
s\rightarrow s-2\bar{\epsilon}_1\Gamma_{\circ}\epsilon_2.
\end{align}


The full action $A=A_{{\rm boson}}+A_{{\rm fermion}}$ now shows that the Gauss constraints corresponding to the $\delta_{HL}$-gauge symmetry are
\begin{align}
&\boldsymbol{G}_{A}\equiv i[\boldsymbol{X}^{\mu}, \boldsymbol{P}_{\mu}]-
\frac{i}{2}[\boldsymbol{\Theta}, \Gamma^0\Gamma_{\circ}\boldsymbol{\Theta}]_+=0, 
\label{gauss1}\\
&\boldsymbol{G}_{B}\equiv X_{\uhdb}\cdot \hat{\boldsymbol{P}}=0, 
\label{gauss2}
\end{align}
and the final result for the effective mass square is, in the 
$\boldsymbol{K}=0$ gauge, 
\begin{align}
{\cal M}^2&=N{\rm Tr}(\hat{\boldsymbol{P}}\cdot \hat{\boldsymbol{P}}) 
-\frac{N}{6}\bigl<
[X^{\mu},X^{\nu}, X^{\sigma}], [
X_{\mu},X_{\nu},X_{\sigma}]
+i\frac{N}{2}\bigl<\bar{\Theta}, \Gamma_{\mu\nu}[
X^{\mu},X^{\nu}, \Theta]\bigr>\\
&=N{\rm Tr}\Bigl[
\hat{\boldsymbol{P}}\cdot 
\hat{\boldsymbol{P}}- \frac{1}{2}\bigl(
X_{\uhdb}^2
[\boldsymbol{X}^{\nu},\boldsymbol{X}^{\sigma}][\boldsymbol{X}_{\nu},\boldsymbol{X}_{\sigma}]
-2[X_{\uhdb}\cdot \boldsymbol{X}, \boldsymbol{X}^{\nu}]
[X_{\uhdb}\cdot\boldsymbol{X}, \boldsymbol{X}_{\nu}]\bigr)
\nonumber \\
&\qquad \qquad+i\bar{\boldsymbol{\Theta}}\Gamma_{\mu\nu}
X_{\uhdb}^{\mu}[\boldsymbol{X}^{\nu}, \boldsymbol{\Theta}]
\Bigr]. 
\end{align}
The first line of (\ref{susyalgebra}) is proportional to 
${\cal M}^2$ under the $\delta_Y$-Gauss constraint and the 
$\boldsymbol{K}$-equation of motion in the $\boldsymbol{K}=0$ gauge, respectively, 
\begin{align}
\boldsymbol{G}_{Z}\equiv P_{\circ}\cdot \hat{\boldsymbol{X}}=0, \quad 
P_{\circ}\cdot \hat{\boldsymbol{P}}=0, 
\label{zgauss+gauge}
\end{align}
in addition to the other Gauss constraints. 
As stressed already in the treatment of the bosonic part, the 
mass-shell condition must be understood in conjunction with 
these Gauss constraints. 
The Gauss constraints 
together with the $\boldsymbol{K}$ equations of motion
are themselves 
invariant under the dynamical supersymmetry, 
\begin{align}
\delta_{\epsilon}\boldsymbol{G}_A=0, \quad 
\delta_{\epsilon}\boldsymbol{G}_B=0, 
\quad 
\delta_{\epsilon}\boldsymbol{G}_Z=0, 
\quad 
\delta_{\epsilon}(P_{\circ}\cdot \hat{\boldsymbol{P}})=0. 
\end{align}

On the other hand, 
${\cal M}^2$ itself is  
not super invariant, but the following 
combination which involves gauge fields and corresponds to the total Hamiltonian 
${\cal H}$ of our system is invariant:
\begin{align}
\delta_{\epsilon}\Bigl(\frac{1}{e}{\cal H}\Bigr)= \delta_{\epsilon}\Bigl({\rm Tr}(\boldsymbol{A}\boldsymbol{G}_A-
\hat{\boldsymbol{B}}\boldsymbol{G}_B+\boldsymbol{Z}
\boldsymbol{G}_{Z}
)
-\frac{1}{2N}{\cal M}^2\Bigr)=0, 
\end{align}
since $\delta_{\epsilon}P_{\circ}^{\mu}=0$, as we have already stressed 
before. 
Thus, the supersymmetry of the effective mass square is 
satisfied only after imposing the Gauss constraints ensuring the consistency of our formalism. 
The same can be said concerning the positivity of the 
effective mass square ${\cal M}^2$, since the 
closure of the supersymmetry algebra (\ref{susyalgebra}) is also 
ensured in conjunction with those Gauss constraints.


Finally, we derive the full effective action in the light-front 
gauge using the light-front coordinates on the M-plane introduced in 
section 3.  We have already seen that the 
projection condition reduces to 
\begin{align}
\Gamma^+\boldsymbol{\Theta}=0, 
\label{lfcondfermi}
\end{align}
resulting
\begin{align}
&\frac{1}{2}{\rm Tr}\Bigl(
\bar{\boldsymbol{\Theta}}\Gamma_{\circ}\frac{D\boldsymbol{\Theta}}{D\tau}\Bigr)
=\frac{1}{4}\sqrt{-\frac{P^+_{\circ}}{P_{\circ}^-}}{\rm Tr}\Bigl(\boldsymbol{\Theta}
\frac{D\boldsymbol{\Theta}}{D\tau}\Bigr), \\
&
-ie\frac{1}{4}\langle \bar{\Theta}, 
\Gamma_{\mu\nu}[X^{\mu},X^{\nu},\Theta]\rangle =
-ie\frac{1}{4}\sqrt{-\frac{P^+_{\circ}}{P_{\circ}^-}}\sqrt{X_{{\rm M}}^2}
{\rm Tr}
(\boldsymbol{\Theta}\Gamma_i[\boldsymbol{X}_i, \boldsymbol{\Theta}]).
\end{align}
Then, by rescaling
\begin{align}
\boldsymbol{\Theta}\rightarrow \sqrt{2}\Bigl(-\frac{P^+_{\circ}}{P_{\circ}^-}\Bigr)^{-1/4}\boldsymbol{\Theta}, 
\end{align}
the Hamiltonian constraint is 
\begin{align}
&P_{\circ}^2+{\cal M}_{{\rm lf}}^2\simeq 0\\
&{\cal M}_{{\rm lf}}^2\equiv 
N{\rm Tr}\bigl(
\hat{\boldsymbol{P}}^i\hat{\boldsymbol{P}}^i
-\frac{1}{2}X_{\rm M}^2 [
\boldsymbol{X}_i,\boldsymbol{X}_j][
\boldsymbol{X}_i,\boldsymbol{X}_j]
+i\sqrt{X_{{\rm M}}^2}\boldsymbol{\Theta}
\Gamma_i[\boldsymbol{X}_i, \boldsymbol{\Theta}]
\bigr).
\end{align}
Repeating the same procedure as in the purely bosonic case
 in section 3, we find the full effective action for traceless matrix variables, 
\begin{align}
A_{{\rm lf}}=
\int ds\, \Bigl[{\rm Tr}\Bigl(
\hat{\boldsymbol{P}}_i\frac{D\hat{\boldsymbol{X}}_i}{Ds}
+\frac{1}{2}\boldsymbol{\Theta}
\frac{D\boldsymbol{\Theta}}{Ds}\Bigr)
-\frac{1}{2N}{\cal M}_{{\rm lf}}^2
\Bigr], 
\end{align}
which is the first-order form of the light-front action.\footnote{Note that 
after the equations of motion including fermionic variables are used, we can set $X_{\circ}^+=\frac{P_{\circ}^+}{N}s$ as in the purely bosonic case of section 3. } We note that in this gauge, the light-like limit $P_{\circ}^-=0$ which has been 
excluded by our assumption can be 
included as a limiting case. 

The 
case of spatial foliation is derived similarly, resulting as 
\begin{align}
A_{{\rm spat}}=\int ds \Bigl[{\rm Tr}\Bigl(\hat{\boldsymbol{P}}_i\frac{D\hat{\boldsymbol{X}}_i}{Ds}
+\frac{1}{2}\boldsymbol{\Theta}
\frac{D\boldsymbol{\Theta}}{Ds}\Bigr)
-P_{\circ}^0\Bigr]
\end{align}
with the same condition (\ref{lfcondfermi}) and 
\begin{align}
P_{\circ}^0=
\sqrt{
(P_{\circ}^{10})^2+{\cal M}_{{\rm lf}}^2
}.
\end{align}
The second-order form of this effective action is given 
as in the bosonic case by solving for the bosonic momenta, resulting with fermion potential term in addition to the purely bosonic 
potential term in 
(\ref{tmbosonpotential}).

\section{Concluding remarks}

We have proposed a consistent re-formulation 
of Matrix theory with 11 dimensional Lorentz covariance, 
as an intermediate step toward ultimate 
formulation of M-theory. We have not, 
needless to say,  proved the uniqueness 
of our construction. Possibilities to deform or extend our formulation  
by modifying or relaxing some of the 
symmetry requirements or by adding higher order 
terms for the potential and kinetic terms are not excluded. 
In connection with this, we stress again that our standpoint toward  
covariantized Matrix theory on the basis of the 
DLCQ interpretation for finite $N$ is not based on the 
naive analogies with the structure of supermembrane action, 
which were mentioned in section 1 as 
a heuristic motivation for discretized Nambu bracket. 
For example, from the classical dynamics of supermembranes, there is no immediate analog for the M-variables, being responsible for the scale invariance and covariant projection conditions as well as 
the crucial higher gauge symmetries 
in our model. 

To conclude, we briefly mention some important issues unsolved or untouched in the 
present work. 

\noindent
(1) We have not 
examined whether our covariant 
reformulation of Matrix theory is 
useful for discussing various possible bound states of 
M-theory partons, especially in the limit of infinite $N$. 
It remains to see whether 11 dimensional coordinate matrices 
together with the M-variables can provide any new insight for 
representing various currents and conserved charges if we treat all components of the 
matrices in a manifestly covariant fashion. In particular, one of 
the important problem is how the transverse M5-branes could be 
realized in the present context. 
Possible reformulations 
of various duality relations among those physical 
objects of M-theory also remain to be investigated. 

\noindent
(2) 
The problem of covariant formulation of currents is closely related to 
the problem of background dependence. Our formulation is 
consistent on the completely flat Minkowski background. 
In view of an interesting observation \cite{maldacenaetal} that 
the single transverse M5-brane corresponds to the trivial 
classical vacuum of the so-called pp-wave matrix theory, 
it may be useful to study the possibility of extending the present 
covariant formulation to a deformed covariantized matrix model corresponding to a pp-wave background of supergravity. 

\noindent
(3) 
In general, however, 
it is not at all 
obvious how to deform the theory to curved backgrounds, 
since the theory is intrinsically non-local and satisfies 
novel gauge symmetries. Unlike the light-front 
formulation, the analogy with super membranes does not work either. 
For these reasons, it 
is not straightforward to define energy-momentum tensor 
and other currents in our framework.\footnote{Possible connection to the 
super-embedding approach (see {\it e.g.} \cite{bandos} and 
references therein) 
may here be worthwhile to pursue, 
since such a geometrical approach seems 
useful in clarifying the relation at least with {\it classical} 11-dimensional supergravity. Note however that it is not at all clear how such 
classical structure could be related the 
generation of non-linear gravitational interactions through the quantum effects 
of matrices, as demonstrated for instance in \cite{oy} and references therein. For bridging them, something analogous to the renormalization 
group approach to world-sheet conformal symmetry 
in string theory, is desirable. 
} Most probably,  the 
higher gauge transformations themselves are deformed or 
extended further in the 
presence of non-trivial backgrounds. The problem is also related to 
the fundamental issue of background {\it independence} of Matrix theory, 
which is expected to be resolved only when 
the theory is treated fully quantum mechanically, because the interactions among the actual gravitational degrees of freedom can only 
emerge as loop effects (see \cite{taylor} for a review on this subject).

\noindent
(4) In the present paper, we 
have restricted ourselves essentially to classical theory. Since we have already 
given the whole structure in the setting of first-order canonical formalism, 
it would be relatively straightforward, at least formally, to formulate fully covariant and 
BRST invariant 
quantizations of our theory both in path-integral and 
operator methods. If we adopt the gauge conditions involving 
$\tau$-derivatives of the gauge fields, we can treat them as 
unphysical propagating fields together with the vector matrix fields and 
ghost fields.  That  
would be useful, for example,  
in applying our formalism to study scattering amplitudes and 
correlation functions, in addition to the problems related to the above 
issues, although we have to be very careful about the 
validity of perturbative methods. 

\noindent
(5) A problem of different nature is whether our 
methods can be extended to a covariantization of 
matrix {\it string} theory \cite{dvv}, in the sense of SO(9,1) 
Lorentz symmetry in 10 dimensions with small $g_s$. The matrix string theory can be 
regarded as a different but equally possible matrix
regularization \cite{sy2} of classical membrane theory,  when 
the membranes are  
wrapped around the M-theory circle. It should  
in principle be possible to extend our covariantized Matrix theory 
by suitably reformulating the procedure of 
compactification with windings. 
A difficult task in this direction is to find a way of reformulating 
Virasoro conditions such that they correspond to the Gauss constraints of 
some higher gauge symmetries associated with 
matrix variables, in analogy with 
our higher gauge symmetries. 
It might provide us a new theory of covariant 
second-quantized strings, differing from the 
standard approach of string field theories. 

\noindent
(6) 
Another important issue concerns anti D-particles. 
As our discussion of gauge fixing in section 3 clearly shows,  
the present theory only allows D-particles as observable 
degrees of freedom. This is also 
consistent with the presence of (dynamical) supersymmetry 
which is realized with a precise matching of traceless 
matrix degrees of freedom between bosonic and 
fermionic variables at each mass level. 
If we treat a system in which 
D-branes and anti-D-branes coexist from the viewpoint 
of 10-dimensional open-string theory, supersymmetry must 
be necessarily spontaneously broken \cite{tywgso}, and 
precise matching of degrees of freedom does not hold 
at each mass level, corresponding to a nonlinear 
realization of supersymmetry.\footnote{As regards to  
approaches from the viewpoint of effective world-volume actions, see \cite{asen} and 
references therein. } It is an interesting question whether and how 
 covariant matrix theory with 
both D-particles and anti-D-particles is possible. 
To answer this question satisfactorily requires 
us to treat the size of matrices as a genuine dynamical variable, 
in order to describe creation and annihilation of 
brane-anti-brane pairs as {\it dynamical} processes. That would also improve consistent but 
somewhat ad hoc nature of relating the 
(light-like) momentum and the size of the matrices in the present formulation 
of Matrix theories, by providing some deeper understanding 
on such a relationship. 
In particular, the higher gauge symmetry must be extended to 
include SU($N$)$\times$SU($M$) with varying $N$ and $M$ such that 
only the difference $N-M$ is strictly conserved. In other words, the theory must be formulated ultimately in a 
Fock space with respect to the sizes of matrices in which 
we can go back and forth among different sizes of 
matrices. 
This is a great challenge, 
perhaps forcing us to invent a new 
theoretical framework. For a tentative attempt 
related to this problem, see ref. \cite{ty2}.

\acknowledgments

I am grateful to H. Awata, M. Li, and D. Minic 
for collaboration in our old work which 
motivated this work. I also thank Y. Sekino for 
his comments on a preliminary version of the manuscript. 

The present work is supported in part by Grant-in-Aid for 
Scientific Research (No. 25287049) from the Ministry of 
Educationl, Science, and Culture.

\appendix
\section{Dirac brackets for the center-of-mass part}

We here briefly discuss two versions of Dirac bracket for the center-of-mass 
variables, 
taking account of the second-class constraint (\ref{c-o-m-fermiprimary}) 
for fermionic center-of-mass variables, 
depending upon whether (a) we treat the bosonic orthogonality Gauss constraint 
(\ref{orthogonality1}) as a weak constraint imposed after 
computing brackets, or (b) as a strong constraint taking 
into account (\ref{orthogonality1}) by 
appropriately fixing the $\delta_{w}$-gauge symmetry.

\vspace{0.2cm}
\noindent
(a) 
The Dirac brackets for bosonic variables are 
\begin{align}
\{X^{\mu}_{\circ}, P^{\nu}_{\circ}\}_{{\rm D_a}}&=
\eta^{\mu\nu}, \quad \{P^{\mu}_{\circ}, P^{\nu}_{\circ}\}_{{\rm D_a}}=0\\
\{X^{\mu}_{\circ}, X^{\nu}_{\circ}\}_{{\rm D_a}}&=
\frac{1}{2}\bar{\Theta}\Gamma^{\mu}(P_{\circ}\cdot\Gamma)^{-1}
\Gamma^{\nu}\Theta.
\label{xxdiracb}
\end{align}
Note that in the last equation antisymmetry with respect to exchange 
$\mu\leftrightarrow \nu$ is ensured by $
(\Gamma^0\Gamma^{\mu}\Gamma^{\sigma}\Gamma^{\nu})^{{\rm T}}
=\Gamma^0\Gamma^{\nu}
\Gamma^{\sigma}\Gamma^{\mu}
$. 
The cases involving fermionic variables are 
\begin{align}
\{\Theta_{\alpha}, X^{\mu}_{\circ}\}_{{\rm D_a}}
&=\frac{1}{2}\bigl((P_{\circ}\cdot\Gamma)^{-1}\Gamma^{\mu}\Theta_{\circ}\bigr)_{\alpha}, \quad 
\{\Theta_{\alpha}, P_{\circ}^{\mu}\}_{{\rm D_a}}=0, \\
\{\Theta_{\alpha}, \bar{\Theta}_{\beta}\}_{{\rm D_a}}&=\frac{1}{2}
(P_{\circ}\cdot \Gamma)^{-1}_{\alpha\beta}. 
\end{align}
Those results are non-singular provided $-P_{\circ}^2>0$. Note 
that the center-of-mass coordinates $X_{\circ}^{\mu}$ are 
not independent of the spinor coordinates, as required by the 
consistency with the supersymmetry transformation laws (\ref{rigidsuper}). 

\vspace{0.2cm}
\noindent
(b) As usual constrained Hamiltonian formalism teaches us, we have 
to impose an appropriate gauge condition, in treating the Gauss constraint (\ref{orthogonality1}) strongly associated with the $\delta_{w}$-gauge symmetry.  
Since this gauge symmetry allows us to shift $X_{\circ}^{\mu}$ along the direction of $X_{{\rm M}}^{\mu}$ 
arbitrarily, we can choose the following Lorentz-invariant gauge fixing condition, 
\begin{align}
X_{\circ}\cdot X_{{\rm M}}=0. 
\label{omegagaugefix}
\end{align}
Although we do not claim that this is the most 
convenient gauge choice, let us use this as a 
simple example of canonical treatment. 
Together with (\ref{orthogonality1}),  the M-variable is now manifestly orthogonal to 
the canonical pair of time-like vectors $(X_{\circ}^{\mu}, P_{\circ}^{\mu})$. 
Then we have 
\begin{align}
&\{X_{\circ}\cdot X_{{\rm M}}, P_{\circ}\cdot X_{{\rm M}}\}_{{\rm P}}=X_{{\rm M}}^2
\end{align}
and, hence, modify the bosonic Dirac brackets of case (a) as 
\begin{align}
\{X_{\circ}^{\mu},P_{\circ}^{\nu}\}_{{\rm D_b}}=
\eta^{\mu\nu}-\frac{1}{(X_{{\rm M}})^2}X^{\mu}_{{\rm M}}X^{\nu}_{{\rm M}}
\label{xpdiracb}
\end{align}
with $\{P_{\circ}^{\mu}, P_{\circ}^{\nu}\}_{{\rm D_a}}$ being intact. 
Then we find \begin{align}
&\{X_{\circ}^{\mu}, P_{{\rm M}}^{\nu}\}_{{\rm D_b}}
=-\frac{1}{X_{{\rm M}}^2}X_{{\rm M}}^{\mu}X_{\circ}^{\nu}, \quad 
\{P_{\circ}^{\mu}, P_{{\rm M}}^{\nu}\}_{{\rm D_b}}
=-\frac{1}{X_{{\rm M}}^2}X_{{\rm M}}^{\mu}P_{\circ}^{\nu},
\\
&
\{X_{\circ}^{\mu}, X_{{\rm M}}^{\nu}\}_{{\rm D_b}}=0, \quad 
\{P_{\circ}^{\mu}, X_{{\rm M}}^{\nu}\}_{{\rm D}_b}=0, \\
&\{P_{{\rm M}}^{\mu}, P_{{\rm M}}^{\nu}\}_{{\rm D_b}}
=\frac{1}{X_{{\rm M}}^2}(P_{\circ}^{\mu}X_{\circ}^{\nu}-X_{\circ}^{\mu}P_{\circ}^{\nu}), \\
&\{X_{{\rm M}}^{\mu}, P_{{\rm M}}^{\nu}\}_{{\rm D_b}}=\eta^{\mu\nu}, \quad \{X_{{\rm M}}^{\mu}, X_{{\rm M}}^{\nu}\}_{{\rm D_b}}=0.
\end{align}

As for the Dirac brackets involving fermionic variables including 
(\ref{xxdiracb}), it is sufficient 
to make a replacement $\Gamma^{\mu}\rightarrow \tilde{\Gamma}^{\mu}$ 
with 
\begin{align}
\tilde{\Gamma}^{\mu}=
\Gamma^{\mu}-\frac{1}{X_{{\rm M}}^2}X^{\mu}_{{\rm M}}
(X_{{\rm M}}\cdot \Gamma), 
\label{tildegamma}
\end{align}
satisfying 
\begin{align}
X_{{\rm M}}\cdot \tilde \Gamma=0. 
\label{gammaorthoM}
\end{align}

The supercharge associated with (\ref{rigidsuper}) is
\begin{align}
Q_{\circ}=2P_{\circ}\cdot \Gamma \Theta_{\circ}. 
\end{align}
Using the first version (a) of the Dirac brackets, we have
\begin{align}
&\{Q_{\circ}, X_{\circ}^{\mu}\}_{{\rm D_a}}=-\Gamma^{\mu}\Theta_{\circ},\quad 
\{Q_{\circ}, P_{\circ}^{\mu}\}_{{\rm D_a}}=0, \\
&\{\bar{Q}_{\circ\, \alpha}, \Theta_{\circ\, \beta}\}_{{\rm D_a}}
=\delta_{\alpha\beta},\\
&\{Q_{\alpha}, \bar{Q}_{\beta}\}_{{\rm D_a}}
=2(P_{\circ}\cdot\Gamma)_{\alpha\beta}, 
\end{align}
which are consistent with the transformation laws. 
If we use the second version (b) of the Dirac bracket, 
$\Gamma^{\mu}$ is replaced by $\tilde{\Gamma}^{\mu}$. 
This and similar modification of 
bosonic brackets exhibited in (\ref{xpdiracb}) are due to the fact that the gauge-fixing condition (\ref{omegagaugefix}) is not 
invariant against the kinematical supersymmetry transformation as 
well as bosonic translation symmetry, and 
hence we have to perform compensating $\delta_w$- 
gauge transformations with field-dependent parameters. For example, the 
compensating gauge parameter associated with the 
supersymmetry transformation is $w=-X_{{\rm M}}\cdot \bar{\varepsilon}\Gamma\Theta/X_{{\rm M}}^2$ corresponding to the second term in the right-hand side of (\ref{tildegamma}). 
Finally, using (\ref{gammaorthoM}), it is easy to check that the M-variables are 
inert under the super transformations.

\section{Derivation of dynamical supersymmetry transformations}
\noindent
(1) {\it Transformation laws}

Since our formulation is completely 
covariant under 11 dimensional Lorentz transformations, 
we are free to use an arbitrary Lorentz frame to study supersymmetry. 
A convenient frame for this purpose is such that only non-zero 
component of $P_{\circ}^{\mu}$ is the time component 
$P_{\circ}^0$, assuming a time-like $P_{\circ}^{\mu}$, and that of $X_{\uhdb}^{\mu}$ is $X_{\uhdb}^{10}$. 
By making 
a boost along the 10-th spatial direction in terms of the usual light-front foliation this is alway possible: 
this frame is characterized by 
$P_{\circ}^+=-P_{\circ}^-$ and hence $X_{\uhdb}^+=X_{\uhdb}^-$ 
due to (\ref{lfrelation}). 
In this frame the projection condition for fermionic 
variables become the ordinary light-like condition, 
\begin{align}
(\Gamma_0-\Gamma_{10})\boldsymbol{\Theta}=0, 
\quad 
(\Gamma_0+\Gamma_{10})\epsilon=0.
\label{speprojection}
\end{align}

Now, by re-definining the gauge field $ \boldsymbol{B}$ as
\begin{align}
\boldsymbol{B}\rightarrow \boldsymbol{B}'=\boldsymbol{B}-
\frac{1}{X_{\uhdb}^{10}}\Bigl(\frac{1}{e}
\frac{d\hat{\boldsymbol{X}}^{10}}{d\tau}+i[\boldsymbol{A}, 
\hat{\boldsymbol{X}}^{10}]-\frac{1}{2}
\hat{\boldsymbol{P}}^{10}\Bigr), 
\label{b'}
\end{align}
we can eliminate the 10-th components of 
matrix variables from the Poincar\'{e} integral and 
the quadratic kinetic term, in terms of $\boldsymbol{B}'$. 
Similarly, we can eliminate the 0-th component 
of the coordinate matrix $\boldsymbol{X}^0$ from the 
potential term, 
by redefining the gauge field $\boldsymbol{Z}$
\begin{align}
\boldsymbol{Z}\rightarrow \boldsymbol{Z}'=
\boldsymbol{Z}+\frac{1}{P_{\circ}^{0}}\Bigl(
\frac{1}{e}\frac{d\hat{\boldsymbol{P}}^{0}}{d\tau}
+i[\boldsymbol{A}, \hat{\boldsymbol{P}}^{0}]
+\frac{1}{2}X_{{\rm M}}^2[\boldsymbol{X}_i, [\boldsymbol{X}^0, \boldsymbol{X}_i]]\Bigr)
\label{z'}
\end{align}
with $i$ running over only SO(9) directions transverse to the M-plane.
Furthermore, 
from the definition of the 3-bracket, in this special frame, the potential term does not 
involve $\boldsymbol{X}^{10}$. 
Thus the remaining terms of the bosonic 
part of action 
are now given by
\begin{align}
A_{{\rm boson}}'=\int d\tau\, {\rm Tr}&
\Bigl[-e\hat{\boldsymbol{P}}^{10}
\boldsymbol{B}'X_{\uhdb}^{10}-e\hat{\boldsymbol{X}}^0P_{\circ}^0
\boldsymbol{Z}'+\frac{e}{2}(\hat{\boldsymbol{P}}^0-P_{\circ}^0\boldsymbol{K})^2\nonumber \\
&
+\hat{\boldsymbol{P}}_i\cdot \Bigl(
\frac{d\hat{\boldsymbol{X}}_i}{d\tau}+ie[\boldsymbol{A}, 
\boldsymbol{X}_i]\Bigr)-\frac{e}{2}(\hat{\boldsymbol{P}}_i)^2+\frac{e}{4}X_{\uhdb}^2[\boldsymbol{X}_i, \boldsymbol{X}_j]^2\Bigr]. 
\end{align}
It should be kept in mind that we 
dropped the part involving the center-of-mass variables and the term   $P_{\dhd}\cdot \frac{dX_{\uhdb}}{d\tau}$ for the M-variables, since under the conservation laws 
of $P_{\circ}^{\mu}$ and $X_{{\rm M}}^{\mu}$ these part of the action behaves as a total derivative. As emphasized in the text, we treat 
this reduced action together with the equations of motion for these variables, with 
the Gauss constraint $P_{\circ}\cdot X_{{\rm M}}=0$ being 
strongly imposed. 
Apart from the terms in the first line, the reduced action shown in the second line is formally 
the same as the  
bosonic part of the action for the ordinary supersymmetric 
quantum mechanics expressed in the first-order formalism. 
Note however that the 11 dimensional covariance is not at all lost in this process: using covariant language, 
non-covariant looking expressions should be 
understood, together with (\ref{orthogonality1}) and (\ref{orthogonality2}), as
\begin{align}
\hat{\boldsymbol{X}}^{10}=\frac{X_{\uhdb}\cdot \hat{\boldsymbol{X}}}
{\sqrt{X_{\uhdb}^2}}, \quad \hat{\boldsymbol{P}}^{10}=\frac{X_{\uhdb}\cdot \hat{\boldsymbol{P}}}{\sqrt{X_{\uhdb}^2}}, 
\quad \hat{\boldsymbol{P}}^0=\frac{P_{\circ}\cdot \hat{\boldsymbol{P}}}
{\sqrt{-P_{\circ}^2}} , \quad 
P_{\circ}^0=\sqrt{-P_{\circ}^2}, \quad 
X_{{\rm M}}^{10}=\sqrt{X_{{\rm M}}^2}
\end{align}
and the index $i$ labels nine independent traceless coordinate 
matrices in an arbitrary (ortho-normal) basis satisfying covariant orthogonality conditions,  
\begin{align}
P_{\circ}\cdot \hat{\boldsymbol{X}}=0, \,\, P_{\circ}\cdot \hat{\boldsymbol{P}}=0, \quad 
X_{\uhdb}\cdot \hat{\boldsymbol{X}}=0, \, \, X_{\uhdb}\cdot \hat{\boldsymbol{P}}=0. \nonumber 
\end{align}
  
We now study the fermionic part of the action on the basis of 
the requirement of supersymmetry, with 
understanding that all the fermion variables below are 
projected as discussed in section 4. Since we are using 
somewhat unfamiliar first-order formalism, we start from scratch. 
First, the kinetic term is chosen to be 
\begin{align}
\frac{1}{2}{\rm Tr}\Bigl(
\bar{\boldsymbol{\Theta}}\Gamma_{\circ}\frac{
D\boldsymbol{\Theta}}{D\tau}\Bigr)=-\frac{1}{2}{\rm Tr}\Bigl(
\frac{D\bar{\boldsymbol{\Theta}}}{D\tau}\Gamma_{\circ}
\boldsymbol{\Theta}\Bigr)
\end{align}
Comparing with the center-of-mass case, this amounts to a change of the normalization of traceless part by $\boldsymbol{\Theta}
\rightarrow (-P_{\circ}^2)^{-1/4}\boldsymbol{\Theta}$ and 
$\epsilon\rightarrow (-P_{\circ}^2)^{1/4}\epsilon$.
Note that this changes the scaling dimensions of 
$\boldsymbol{\Theta}$ and $\epsilon$ to zero and $1$, respectively. 
The change due to the transformation 
\begin{align}
\delta_{\epsilon}\hat{\boldsymbol{X}}_i=
\bar{\epsilon}\Gamma_i\boldsymbol{\Theta}=-
\bar{\boldsymbol{\Theta}}\Gamma_i\epsilon
\label{1stbosontrans}
\end{align}
in the Poincar\'{e} integral is then cancelled by that of the
 fermionic kinetic term with 
\begin{align}
\delta_{\epsilon}^{(1)}\boldsymbol{\Theta}=\Gamma_{\circ}
\Gamma_i\hat{\boldsymbol{P}}_i\epsilon, \quad 
\delta_{\epsilon}^{(1)}\bar{\boldsymbol{\Theta}}=\hat{\boldsymbol{P}}_i\bar{\epsilon}\Gamma_i\Gamma_{\circ}, 
\quad \delta_{\epsilon}^{(1)}\hat{\boldsymbol{P}}_i=0. 
\label{firstorderfermisusy}
\end{align}
We note that together with 
\begin{align}
\delta_{\epsilon}\hat{\boldsymbol{X}}^0=0, \quad 
\delta_{\epsilon}\hat{\boldsymbol{X}}^{10}=0, \quad 
\delta_{\epsilon}^{(1)}\boldsymbol{P}^{10}=0, \quad 
\delta_{\epsilon}^{(1)}\boldsymbol{P}^0=0, 
\end{align}
these transformations can be brought into covariant form, due to the projection 
(\ref{speprojection}), 
as discussed in the text, namely 
\begin{align}
\delta_{\epsilon}\boldsymbol{X}^{\mu}=\bar{\epsilon}\Gamma^{\mu}
\boldsymbol{\Theta}=-\bar{\boldsymbol{\Theta}}\Gamma^{\mu}\epsilon. 
\end{align}
Similarly, the transformation (\ref{firstorderfermisusy}) is 
covariantized by expressing it as 
\begin{align}
\delta^{(1)}_{\epsilon}\boldsymbol{\Theta}=P_-
\Gamma_{\circ}\Gamma_{\mu}\hat{\boldsymbol{P}}^{\mu}\epsilon, 
\quad \delta^{(1)}_{\epsilon}\bar{\boldsymbol{\Theta}}=\hat{\boldsymbol{P}}^{\mu}
\bar{\epsilon}\Gamma_{\mu}\Gamma_{\circ}P_+
, \quad 
\delta_{\epsilon}^{(1)}\hat{\boldsymbol{P}}^{\mu}=0.
\label{fermitrans1st}
\end{align}

Then, in order to cancel the effect due to (\ref{1stbosontrans}) 
on the potential term, 
we add a corresponding fermionic potential term
\begin{align}
i\frac{\sqrt{X_{{\rm M}}^2}}{2}e{\rm Tr}\bigl(
\bar{\boldsymbol{\Theta}}\Gamma^0\Gamma_i[\boldsymbol{X}_i, 
\boldsymbol{\Theta}]\bigr)
&=-i\frac{e\sqrt{X_{{\rm M}}^2}}{2}{\rm Tr}\bigl(
\bar{\boldsymbol{\Theta}}\Gamma_{\circ}\Gamma_i[\boldsymbol{X}_i, 
\boldsymbol{\Theta}]\bigr)
=-i\frac{e\sqrt{X_{{\rm M}}^2}}{2}{\rm Tr}\bigl(
\bar{\boldsymbol{\Theta}}\Gamma_{\uhdb}\Gamma_i[\boldsymbol{X}_i, 
\boldsymbol{\Theta}]\bigr)\nonumber \\
&= -i\frac{e\sqrt{X_{{\rm M}}^2}}{2}\frac{1}{\sqrt{X_{\uhdb}^2}}X^{10}_{\uhdb}{\rm Tr}
\bigl(\bar{\boldsymbol{\Theta}} \Gamma_{10}\Gamma_i
[\boldsymbol{X}_i, \boldsymbol{\Theta}]
\bigr)
\nonumber \\
&=-
\frac{i}{4}e
\bigr<\bar{\Theta}, \Gamma_{\mu\nu}[X^{\mu},X^{\nu}, \Theta]\bigr>.
\end{align}
Note that 
due to the fermion projection condition and 
the 
structure of the 3-bracket, neither the time nor 10-th spatial components  
of the bosonic traceless matrix contribute in the 
covariantized expression given in the last line. 
Under the fermion transformation (\ref{fermitrans1st}), we have
\begin{align}
-\delta^{(1)}_{\epsilon}\, \Bigl(i\frac{e\sqrt{X_{{\rm M}}^2}}{2}{\rm Tr}\bigl(
\bar{\boldsymbol{\Theta}}\Gamma_{\circ}\Gamma_i[\boldsymbol{X}_i, 
\boldsymbol{\Theta}]\bigr)\Bigr)&=-
ie\sqrt{X_{{\rm M}}^2}{\rm Tr}\bigl(
\bar{\boldsymbol{\Theta}}\Gamma_i\Gamma_j \epsilon[\boldsymbol{X}_i, \boldsymbol{P}_j]\bigr)\nonumber \\
=-ie\sqrt{X_{{\rm M}}^2}{\rm Tr}\bigl(
\bar{\boldsymbol{\Theta}}\epsilon&
[\boldsymbol{X}_i, \boldsymbol{P}_i]\bigr)
-ie \sqrt{X_{{\rm M}}^2}{\rm Tr}\bigl(
\bar{\boldsymbol{\Theta}}\Gamma_{ij}\epsilon
[\boldsymbol{X}_i, \boldsymbol{P}_j]\bigr).
\end{align}
The first term is canceled by transforming the gauge field $\boldsymbol{A}$ in the bosonic term of the Poincar\'{e} integral as
\begin{align}
\delta_{\epsilon}^{(2)}\boldsymbol{A}=\sqrt{X_{{\rm M}}^2}\bar{\boldsymbol{\Theta}}
\epsilon, 
\label{gaugesusy}
\end{align}
which is, 
being a Lorentz scalar, already of covariant form, 
while the second term is done by the bosonic (quadratic) kinetic term if 
 the bosonic momenta transform as
\begin{align}
\delta^{(2)}_{\epsilon}\hat{\boldsymbol{P}}_j=-i\sqrt{X_{{\rm M}}^2}[\bar{\boldsymbol{\Theta}}
\Gamma_{ij}\epsilon, \boldsymbol{X}_i]. 
\label{momentum2ndorder}
\end{align}
This result, being supplemented by 
\begin{align}
\delta^{(2)}_{\epsilon}\hat{\boldsymbol{P}}^0=0, \quad
\delta^{(2)}_{\epsilon}\hat{\boldsymbol{P}}^{10}=0, \quad 
\delta^{(2)}_{\epsilon}\boldsymbol{K}=0, 
\label{zerotrans}
\end{align}
can be covariantized, due to the fermion projection and 
the strong constraint $P_{\circ}\cdot 
X_{\uhdb}=0$, as 
\begin{align}
\delta^{(2)}_{\epsilon}\hat{\boldsymbol{P}}_{\nu}=
-i\sqrt{X_{{\rm M}}^2}[\bar{\boldsymbol{\Theta}}\Gamma_{\mu\nu}\epsilon, \tilde{\boldsymbol{X}}^{\mu}]=i\sqrt{X_{{\rm M}}^2}[
\bar{\epsilon}\Gamma_{\mu\nu}\boldsymbol{\Theta}, \tilde{\boldsymbol{X}}^{\mu}]
\end{align}
satisfying
\begin{align}
P_{\circ}\cdot \delta^{(2)}_{\epsilon}\hat{\boldsymbol{P}}=0, 
\quad 
X_{{\rm M}}\cdot \delta^{(2)}_{\epsilon}\hat{\boldsymbol{P}}=0.
\end{align}
We have here defined 
\begin{align}
\tilde{\boldsymbol{X}}^{\mu}=
\boldsymbol{X}^{\mu}-
\frac{1}{X_{\uhdb}^2}X_{\uhdb}^{\mu}
(\boldsymbol{X}\cdot X_{\uhdb} )
-\frac{1}{P_{\circ}^2}P_{\circ}^{\mu}(\boldsymbol{X}\cdot 
P_{\circ}), 
\end{align}
which is orthogonal to the M-plane. 

Now, this forces us to study the effect of the new contribution (\ref{momentum2ndorder}) on 
the Poincar\'{e} integral:
\begin{align}
\int d\tau\, {\rm Tr}\Bigl(\delta^{(2)}_{\epsilon}\hat{\boldsymbol{P}}_j
\frac{D\hat{\boldsymbol{X}}_j}{D\tau}\Bigr)
&=-i\int d\tau\, \sqrt{X_{{\rm M}}^2}{\rm Tr}\Bigl(
\bar{\boldsymbol{\Theta}}\Gamma_{ij}\epsilon 
\Bigl[\boldsymbol{X}_i, \frac{D\boldsymbol{X}_j}{D\tau}\Bigr]
\Bigr)
\nonumber \\
&=i\frac{1}{2}\int d\tau \, \sqrt{X_{{\rm M}}^2}{\rm Tr}\Bigl(
\frac{D\bar{\boldsymbol{\Theta}}}{D\tau}\Gamma_{ij}\epsilon 
[\boldsymbol{X}_i, \boldsymbol{X}_j]
\Bigr). 
\label{psecond}
\end{align}
This result cancels against the contribution of fermion kinetic term by correcting the transformation of 
fermion matrices, 
\begin{align}
&\delta^{(2)}_{\epsilon}\boldsymbol{\Theta}=-
i\frac{\sqrt{X_{{\rm M}}^2}}{2}\Gamma_{\circ}\Gamma_{ij}\epsilon [\boldsymbol{X}_i, \boldsymbol{X}_j]=-
i\frac{\sqrt{X_{{\rm M}}^2}}{2}P_-\Gamma_{\circ}\Gamma^{\mu\nu}\epsilon [\tilde{\boldsymbol{X}}_{\mu}, \tilde{\boldsymbol{X}}_{\nu}], \\
&\delta^{(2)}_{\epsilon}\bar{\boldsymbol{\Theta}}=-
i\frac{\sqrt{X_{{\rm M}}^2}}{2}[\tilde{\boldsymbol{X}}_{\mu}, \tilde{\boldsymbol{X}}_{\nu}]\bar{\epsilon}\Gamma^{\mu\nu}\Gamma_{\circ}P_+. 
\end{align}
Thus we have a further new contribution from the variation of the 
fermionic potential term
\begin{align}
-\delta^{(2)}_{\epsilon} \Bigl(i\frac{\sqrt{X_{{\rm M}}^2}}{2}e{\rm Tr}\bigl(
\bar{\boldsymbol{\Theta}}\Gamma_{\circ}\Gamma_i[\boldsymbol{X}_i, 
\boldsymbol{\Theta}]\bigr)\Bigr)&=-\frac{X_{{\rm M}}^2}{2}e
{\rm Tr}\bigl(
\bar{\boldsymbol{\Theta}}\Gamma_{\circ}\Gamma_i[\boldsymbol{X}_i, \Gamma_{\circ}\Gamma_{jk}\epsilon [\boldsymbol{X}_j, \boldsymbol{X}_k]]
\bigr)\nonumber \\
&=eX_{{\rm M}}^2
{\rm Tr}\bigl([\boldsymbol{X}_i, \boldsymbol{X}_j]
[\boldsymbol{X}_i, \bar{\boldsymbol{\Theta}}\Gamma_j\epsilon]\bigr)
\label{varfermionpotential}
\end{align}
which is canceled by the contribution from the 
bosonic potential term, with 
\begin{align}
\delta_{\epsilon} \Bigl(e\frac{X_{\uhdb}^2}{4}{\rm Tr}
\bigl([\boldsymbol{X}_i, \boldsymbol{X}_j]^2
\bigr)\Bigr)
=-eX_{\uhdb}^2
{\rm Tr}\bigl(
[\boldsymbol{X}_i, \boldsymbol{X}_j]
[\boldsymbol{X}_i, \bar{\boldsymbol{\Theta}}\Gamma_j\epsilon]\bigr). 
\end{align}
It is to be noted that in deriving (\ref{varfermionpotential}) 
use was made of the Jacobi identity, which amounts 
to the Fundamental Identity (\ref{FI}) 
in terms of the original 3-bracket notation for the potential terms. 

There remain the contributions of 3rd-order with respect to the 
fermion matrices, one of which is the fermionic potential term 
substituted by $\delta_{\epsilon}\hat{\boldsymbol{X}}_i$ and another comes 
from the fermion kinetic term substituted 
by (\ref{gaugesusy}).  The cancellation of the sum of these two terms is ensured by a 
well-known identity for the 11 dimensional  Dirac matrices,
\begin{align}
{\rm Tr}\bigl(\boldsymbol{\Theta}_a\{
\boldsymbol{\Theta}_b,\boldsymbol{\Theta}_c\}\bigr)
\epsilon_d(\Gamma^{\mu})_{(ac}(\Gamma^0\Gamma_{\mu})_{bd)}=0
\end{align}
where the spinor indices are totally symmetrized. 
Taking into account the projection conditions and the symmetry 
properties of the 11 dimensional Dirac matrices in the 
Majorana representation, 
this identity can be reduced to 
\begin{align}
\frac{1}{2}{\rm Tr}\bigl(\boldsymbol{\Theta}_a\{
\boldsymbol{\Theta}_b,\boldsymbol{\Theta}_c\}\bigr)
\epsilon_d
(\Gamma^{i})_{ac}(\Gamma^0\Gamma_{i})_{bd}+\frac{1}{2}{\rm Tr}\bigl(\boldsymbol{\Theta}_a\{
\boldsymbol{\Theta}_b,\boldsymbol{\Theta}_c\}\bigr)
\epsilon_d(\Gamma^{0})_{ad}\delta_{bc}=0, 
\end{align}
in which the first and the second term on the left-hand side correspond, 
respectively, to the former and latter contributions of 3rd order. 

Now we have to go back to the redefinitions, (\ref{b'}) and (\ref{z'}).  
Since our derivation was made under the presumption 
$\delta_{\epsilon}\boldsymbol{B}'=0$ and $\delta_{\epsilon}
\boldsymbol{Z}'=0$, the transformations of the 
gauge fields $\boldsymbol{B}$ and $\boldsymbol{Z}$ are 
determined as
\begin{align}
&\delta_{\epsilon}\boldsymbol{B}=i(X_{\uhdb}^2)^{-1}[\delta_{\epsilon}\boldsymbol{A}, 
X_{\uhdb}\cdot 
\boldsymbol{X}], 
\label{bsusy}\\
&\delta_{\epsilon}\boldsymbol{Z}=i(P_{\circ}^2)^{-1}[\delta_{\epsilon}\boldsymbol{A}, P_{\circ}\cdot \boldsymbol{P}]
+\frac{X_{{\rm M}}^2}{2P_{\circ}^2}
([\delta_{\epsilon}\boldsymbol{X}^{\mu}, [P_{\circ}\cdot \boldsymbol{X}, 
\boldsymbol{X}_{\mu}]]+
[\boldsymbol{X}^{\mu}, [P_{\circ}\cdot \boldsymbol{X}, 
\delta_{\epsilon}\boldsymbol{X}_{\mu}]]).
\label{zsusy}
\end{align}

We have thus established that the reduced action $A'_{{\rm boson}}+A'_{{\rm fermi}}$ is invariant under 
the following covariant dynamical supersymmetry transformations. 
\begin{align}
&\delta_{\epsilon}\hat{\boldsymbol{X}}^{\mu}=\bar{\epsilon}\Gamma^{\mu}
\boldsymbol{\Theta}=-\bar{\boldsymbol{\Theta}}\Gamma^{\mu}\epsilon, \\
&\delta_{\epsilon}\hat{\boldsymbol{P}}_{\mu}=
\delta_{\epsilon}^{(2)}\hat{\boldsymbol{P}}_{\mu}, \\
&\delta_{\epsilon}\boldsymbol{\Theta}=
\delta_{\epsilon}^{(1)}\boldsymbol{\Theta}+\delta_{\epsilon}^{(2)}
\boldsymbol{\Theta}, 
\quad 
\delta_{\epsilon}\boldsymbol{A}=\delta_{\epsilon}^{(2)}\boldsymbol{A}, 
\end{align}
adjoined with (\ref{bsusy}), (\ref{zsusy}) and 
\begin{align}
\delta_{\epsilon}\boldsymbol{K}=0, \quad 
\delta_{\epsilon}P_{\circ}^{\mu}=0, \quad 
\delta_{\epsilon}X_{{\rm M}}^{\mu}=0.
\end{align}
It is to be noted, as one of the merits of 
our first-order formalism, that 
the supersymmetry is actually valid for the derivative part and 
the remaining part proportional to the ein-bein $e$ separately, corresponding to the 
fact that the transformation laws themselves do not 
involve $e$ explicitly. 

The super transformation corresponds to the 
supercharge 
\begin{align}
{\cal Q}=P_-{\rm Tr}(\tilde{\hat{\boldsymbol{P}}}_{\mu}\Gamma^{\mu}\boldsymbol{\Theta}
-\frac{i}{2}\sqrt{X_{{\rm M}}^2}[\tilde{\boldsymbol{X}}^{\mu}, \tilde{\boldsymbol{X}}^{\nu}]\Gamma_{\mu\nu}\boldsymbol{\Theta})
\end{align}
with 
\begin{align}
\tilde{\hat{\boldsymbol{P}}}^{\mu}=
\hat{\boldsymbol{P}}^{\mu}-
\frac{1}{X_{\uhdb}^2}X_{\uhdb}^{\mu}
(\hat{\boldsymbol{P}}\cdot X_{\uhdb} )
-\frac{1}{P_{\circ}^2}P_{\circ}^{\mu}(\hat{\boldsymbol{P}}\cdot 
P_{\circ}).
\end{align}

\vspace{0.2cm}
\noindent
(2) {\it Super transformations of passive variables}

As we have stressed, the super transformation laws derived 
above are valid under the 
conservation laws of $P_{\circ}^{\mu}$ and $X_{{\rm M}}^{\mu}$ with 
the Gauss constraint (\ref{orthogonality1}) being 
imposed strongly.  The cyclic variabels $X_{\circ}^{\mu}$ and
 $P_{{\rm M}}^{\mu}$ 
corresponding to them 
are passively determined by integrating their first-order equations of motion.  The following general argument shows that 
transformation laws for them can also be 
expressed locally in terms of the variation of supercharges 
with respect to their conjugate variables, 
{\it on-shell} after using the equations of motion and 
the Gauss (and associated gauge fixing) constraints for other non-cyclic variables 
including matrix variables.  
The equations of motion for cyclic variables $O$ in general take 
the form
\begin{align}
\frac{dO}{d\tau}
=-\frac{\partial {\cal H}}{\partial \tilde{O}}\equiv \{O, {\cal H}\}_{{\rm D}}
\label{passiveeqom}
\end{align}
with $(O, \tilde{O})$ being a canonical pair of cyclic variables up to a sign, and 
${\cal H}$ is the Hamiltonian, the part of the Lagrangian which is 
proportional to the ein-bein $e$ and, by the definition 
of the cyclic variables, does not 
involve $O$. For example, for $O=X_{\circ}^{\mu}$, 
$\tilde{O}=P_{\circ}^{\mu}$ and for $O=P_{{\rm M}}^{\mu}$, 
$\tilde{O}=-X_{{\rm M}}^{\mu}$. Though we used notations with 
ordinary derivatives for the purpose of 
making the concepts clear, it should be kept in mind that the derivatives with respect to 
canonical variables here and in what follows are to be defined through 
appropriate Dirac bracket as indicated by the equality $\equiv$. 
Under the Gauss constraints which are 
themselves invariant under supersymmetry 
transformations, the supersymmetry of the action 
is equivalent to 
\begin{align}
\{\bar{\epsilon}{\cal Q}, {\cal H}\}_{{\rm D}}=0.
\end{align}
This leads to 
\begin{align}
\{\bar{\epsilon}\frac{\partial {\cal Q}}{\partial \tilde{O}}, 
{\cal H}\}_{{\rm D}}+\{\bar{\epsilon}{\cal Q}, \frac{\partial {\cal H}}{\partial \tilde{O}}\}
_{{\rm D}}=0.
\end{align}
Since ${\cal H}$ is completely independent of 
$O$,  the super transformation of (\ref{passiveeqom}) directly gives 
the time derivative of the super transformation of passive variable as
\begin{align}
\frac{d \delta_{\epsilon}O}{d\tau}=
\{\bar{\epsilon}{\cal Q}, \frac{\partial {\cal H}}{\partial \tilde{O}}\}_{{\rm D}}=
-\{\bar{\epsilon}\frac{\partial {\cal Q}}{\partial \tilde{O}}, 
{\cal H}\}_{{\rm D}}
\end{align}
which equals 
$-\frac{d}{d\tau}\Bigl(\bar{\epsilon}\frac{\partial {\cal Q}}{\partial \tilde{O}}
\Bigr)$ by the equations of motion. 
Then we can set
\begin{align}
\delta_{\epsilon}O=-\bar{\epsilon}\frac{\partial {\cal Q}}{\partial \tilde{O}}
\equiv \bar{\epsilon}\{O, \cal{Q}\}_{{\rm D}}.
\end{align}
For the validity of this argument, we have to define the Dirac bracket 
appropriately by taking into account the gauge-fixing conditions in order to treat the 
Gauss constraints strongly, as already alluded to above. In the present paper, we do not 
elaborate further along this line. Concerning this and other aspects 
of supersymmetry, 
there might be better and technically more elegant 
formulations.

\vspace{0.2cm}
\noindent
(3) {\it The equations of motion for matrix variables}

As an example of  checking the dynamical super transformation laws, 
let us 
confirm consistency with the equations of motion in a simplest case: 
the first order equation of motion for bosonic variables, including 
the Gauss constraint of the $\delta_Y$-gauge symmetry, 
\begin{align}
&\hat{\boldsymbol{P}}^{\mu}-P_{\circ}^{\mu}\boldsymbol{K}
=
\frac{1}{e}\frac{d\hat{\boldsymbol{X}}^{\mu}}{d\tau}
+i[\boldsymbol{A},\boldsymbol{X}^{\mu}]
-\boldsymbol{B}X_{\uhdb}^{\mu}, \label{1steomforP}\\
&P_{\circ}\cdot (\hat{\boldsymbol{P}}-P_{\circ}\boldsymbol{K})=0, \\
&P_{\circ}\cdot \hat{\boldsymbol{X}}=0,\label{zgauss}
\end{align}
we find that 
the transformation of the right-hand side of (\ref{1steomforP}) is 
\begin{align}
-\frac{1}{e}\frac{D\bar{\boldsymbol{\Theta}}}{D\tau}
\Gamma^{\mu}\epsilon
+i\sqrt{X_{\uhdb}^2}\, \bigl([\bar{\boldsymbol{\Theta}}, \boldsymbol{X}^{\mu}]-i(X_{{\rm M}}^2)^{-1}[\bar{\boldsymbol{\Theta}}, X_{\uhdb}\cdot \boldsymbol{X}]X_{\uhdb}^{\mu}\bigr)\epsilon. 
\label{pxrelation}
\end{align}
Using the following equality, being valid under 
(\ref{zgauss}), for the second term
\begin{align}
&-ie[\bar{\boldsymbol{\Theta}}\Gamma_{\circ}\Gamma^{\nu},\boldsymbol{X}_{\nu}
-\frac{1}{X_{\uhdb}^2}X_{\uhdb\, \nu}
(X_{\uhdb}\cdot \boldsymbol{X})]\Gamma_{\circ}\Gamma^{\mu}
\epsilon\nonumber \\
&=-ie[\bar{\boldsymbol{\Theta}}\epsilon, \boldsymbol{X}^{\mu}
-(X_{\uhdb}^2)^{-1}X_{\uhdb}^{\mu}
(X_{\uhdb}\cdot \boldsymbol{X})]
-ie [\bar{\boldsymbol{\Theta}}\Gamma^{\nu\mu}\epsilon, \boldsymbol{X}_{\nu}
-(X_{\uhdb}^2)^{-1}X_{\uhdb\, \nu}
(X_{\uhdb}\cdot \boldsymbol{X})], \nonumber \\
&=-ie[\bar{\boldsymbol{\Theta}}\epsilon, \boldsymbol{X}^{\mu}
-(X_{\uhdb}^2)^{-1}X_{\uhdb}^{\mu}
(X_{\uhdb}\cdot \boldsymbol{X})]
-ie [\bar{\boldsymbol{\Theta}}\Gamma^{\nu\mu}\epsilon, \tilde{\boldsymbol{X}}_{\nu}]\end{align}
and the fermionic equation of motion
\begin{align}
\frac{D\bar{\boldsymbol{\Theta}}}{D\tau}\Gamma_{\circ}
+ie\sqrt{X_{{\rm M}}^2}[\bar{\boldsymbol{\Theta}}\Gamma_{\circ}\Gamma_{\mu}, 
\boldsymbol{X}^{\mu}
-(X_{\uhdb}^2)^{-1}X_{\uhdb}^{\mu}
(X_{\uhdb}\cdot \boldsymbol{X})]P_-=0, 
\end{align}
 we find that 
the right-hand side of (\ref{pxrelation}) is equal to $\delta_{\epsilon}\hat{\boldsymbol{P}}^{\mu}$. 

Up to this point, it was not necessary to use the 
equation of motion for the bosonic mometum matrices, 
\begin{align}
\frac{D\hat{\boldsymbol{P}}_{\mu}}{D\tau}=e
\bigl(&X_{\uhdb}^2[\boldsymbol{X}^{\nu}, [\boldsymbol{X}_{\mu}, 
\boldsymbol{X}_{\nu}]]
-X_{\uhdb\, \mu}[\boldsymbol{X}^{\nu},[X_{\uhdb}\cdot 
\boldsymbol{X}, \boldsymbol{X}_{\nu}]]
-[X_{\uhdb}\cdot \boldsymbol{X}, [\boldsymbol{X}_{\mu}, 
X_{\uhdb}\cdot \boldsymbol{X}]]\bigr)\nonumber \\
&+\frac{ie}{2}X_{\uhdb}^{\nu}[\bar{\boldsymbol{\Theta}}, \Gamma_{\nu\mu}
\boldsymbol{\Theta}]_+
\end{align}
where the symbol $[\quad, \quad]_+$ means anti-commutator 
with respect to matrices. 
Although this case is somewhat more cumbersome than above, we can 
check similarly that the super-transformations of both sides matches on using the fermion equations of motion. 
The simplest way of doing this is to  
use the special frame 
introduced in (1).  


\end{document}